\documentclass[12pt]{article}
\usepackage{amsmath}
\usepackage{hyperref} 
\numberwithin{equation}{section}

\def\Tr{{\rm Tr}}

\def\K3{\mathrm K3}

\def\double #1{#1{\hbox{\kern-2pt $#1$}}}
\def\p {\partial}
\def\half {\frac{1}{2}}

\def\la{\lambda}

\def\ad{{\dot{\alpha}}}
\def\bd{{\dot{\beta}}}
\def\nn{{\nonumber}}
\newcommand{\beq}{\begin{equation}}
\newcommand{\eeq}{\end{equation}}
\newcommand{\bea}{\begin{eqnarray}}
\newcommand{\eea}{\end{eqnarray}}
\def\p{{\partial}}

\usepackage{enumerate}
\usepackage{amssymb}

\def \p{\partial}
\def \a{\alpha}
\def \b{\beta}
\def \e{\epsilon}
\def \d{\delta}
\def \m{\mu}
\def \n{\nu}

\def \l{\lambda}
\def \L{\Lambda}
\def \hl{\hat\lambda}
\def \o{\omega}
\def \ho{\hat\omega}
\def \ha{{\hat\a}}
\def \hb{{\hat\b}}
\def \hm{{\hat\m}}
\def \hn{{\hat\n}}
\def \hA{{\hat A}}
\def \hB{{\hat B}}
\def \hC{{\hat C}}
\def \hD{{\hat D}}
\def \P{\nabla}

\def\pp{{\mathchoice
          {
              \kern 1pt%
              \raise 1pt
              \vbox{\hrule width5pt height0.4pt depth0pt
                    \kern -2pt
                    \hbox{\kern 2.3pt
                          \vrule width0.4pt height6pt depth0pt
                          }
                    \kern -2pt
                    \hrule width5pt height0.4pt depth0pt}%
                    \kern 1pt
           }
            {
              \kern 1pt%
              \raise 1pt
              \vbox{\hrule width4.3pt height0.4pt depth0pt
                    \kern -1.8pt
                    \hbox{\kern 1.95pt
                          \vrule width0.4pt height5.4pt depth0pt
                          }
                    \kern -1.8pt
                    \hrule width4.3pt height0.4pt depth0pt}%
                    \kern 1pt
            }
            {
              \kern 0.5pt%
              \raise 1pt
              \vbox{\hrule width4.0pt height0.3pt depth0pt
                    \kern -1.9pt  
                    \hbox{\kern 1.85pt
                          \vrule width0.3pt height5.7pt depth0pt
                          }
                    \kern -1.9pt
                    \hrule width4.0pt height0.3pt depth0pt}%
                    \kern 0.5pt
            }
            {
              \kern 0.5pt%
              \raise 1pt
              \vbox{\hrule width3.6pt height0.3pt depth0pt
                    \kern -1.5pt
                    \hbox{\kern 1.65pt
                          \vrule width0.3pt height4.5pt depth0pt
                          }
                    \kern -1.5pt
                    \hrule width3.6pt height0.3pt depth0pt}%
                    \kern 0.5pt
            }
        }}

\def\mm{{\mathchoice
                       {
                             \kern 1pt
               \raise 1pt    \vbox{\hrule width5pt height0.4pt depth0pt
                                  \kern 2pt
                                  \hrule width5pt height0.4pt depth0pt}
                             \kern 1pt}
                       {
                            \kern 1pt
               \raise 1pt \vbox{\hrule width4.3pt height0.4pt depth0pt
                                  \kern 1.8pt
                                  \hrule width4.3pt height0.4pt depth0pt}
                             \kern 1pt}
                       {
                            \kern 0.5pt
               \raise 1pt
                            \vbox{\hrule width4.0pt height0.3pt depth0pt
                                  \kern 1.9pt
                                  \hrule width4.0pt height0.3pt depth0pt}
                            \kern 1pt}
                       {
                           \kern 0.5pt
             \raise 1pt  \vbox{\hrule width3.6pt height0.3pt depth0pt
                                  \kern 1.5pt
                                  \hrule width3.6pt height0.3pt depth0pt}
                           \kern 0.5pt}
                       }}

\def\ad{{\kern0.5pt
                   \alpha \kern-5.05pt
\raise5.8pt\hbox{$\textstyle.$}\kern 0.5pt}}

\def\bd{{\kern0.5pt
                   \beta \kern-5.05pt \raise5.8pt\hbox{$\textstyle.$}\kern 0.5pt}}

\def\qd{{\kern0.5pt
                   q \kern-5.05pt \raise5.8pt\hbox{$\textstyle.$}\kern 0.5pt}}
\def\Dot#1{{\kern0.5pt
     {#1} \kern-5.05pt \raise5.8pt\hbox{$\textstyle.$}\kern 0.5pt}}

\begin{document}
\setcounter{page}0
\thispagestyle{empty}
\begin{flushright}
\makebox[0pt][b]{HU-EP-15/41}
\end{flushright}
\vspace{30pt}

\begin{center}
{\LARGE Worldsheet dilatation operator for the $AdS$ superstring}\\

\vspace{30pt}
{Israel Ram\'{\i}rez${}^{\clubsuit\diamondsuit}$\footnote{email: ramirezkrause@gmail.com} and Brenno Carlini
Vallilo${}^{\spadesuit}$\footnote{email: vallilo@gmail.com}
}
\vspace{30pt}

${}^{ \clubsuit}${\em Departamento de F\'isica, Universidad T\'ecnica Federico Santa Mar\'ia,\\
Casilla 110-V, Valpara\'iso, Chile}

 ${}^{ \diamondsuit}${\em Institut f\"ur Mathematik und Institut f\"ur Physik,\\ Humboldt-Universit\"at zu Berlin\\
IRIS Haus, Zum Gro{\ss}en Windkanal 6,  12489 Berlin, Germany}

${}^{ \spadesuit}${\em Departamento de Ciencias F\'{\i}sicas,
Universidad Andres Bello, \\Republica 220, Santiago, Chile}
\vspace{60pt}

 {\bf Abstract}
\end{center}
In this work we propose a systematic way to compute the logarithmic divergences of
composite operators in the pure spinor description of the $AdS_5\times S^5$ superstring.
The computations of these divergences can be summarized in terms of a dilatation operator
acting on the local operators. We check our results with some important composite operators of the formalism.

\newpage

{ \parskip=0.0in \tableofcontents }


\section{Introduction}

During the last decade there was a great improvement in the
understanding of $\mathcal{N}=4$ super Yang-Mills theory due to integrability
techniques,  culminating in a proposal where the anomalous dimension of
any operator can be computed at any coupling \cite{Gromov:2009bc}.  The
crucial point of this advance was the realization that the
computations of anomalous dimensions could be systematically done by
studying the dilatation operator of the theory \cite{Beisert:2003jj,Beisert:2003yb}. For a general review and
an extensive list of references, we recommend \cite{Beisert:2010jr}. An alternative to the TBA approach not covered in
\cite{Beisert:2010jr}, the Quantum Spectral Curve, was developed
in \cite{Gromov:2013pga,Gromov:2014caa}. For some of its applications,
including high loops computations, see
\cite{Gromov:2014bva,Alfimov:2014bwa,Marboe:2014gma,Marboe:2014sya,Gromov:2015wca,Gromov:2015vua}

On the string theory side it is that known the world sheet sigma-model  is
classically integrable \cite{Bena:2003wd,Vallilo:2003nx}. However, it is not yet
known how to fully quantize the theory, identifying all physical vertex operators
and their correlation functions. In the case of the pure spinor string
it is known that the model is conformally invariant at all orders of perturbation
theory and that the non-local charges found in \cite{Vallilo:2003nx} exist
in the quantum theory \cite{Berkovits:2004xu}. In a very interesting paper, \cite{Benichou:2012hc}
showed how to obtain the Y-system equations from the holonomy operator.

Another direction in which the pure spinor formalism was used with success was
the quantization around classical configurations. In \cite{Aisaka:2012ud} it was shown
that the semi-classical quantization of a large class of classical backgrounds agrees
with the Green-Schwarz formalism. This was later generalized in \cite{Cagnazzo:2012uq,Tonin:2013uec}.
Previously, Mazzucato and one of the authors \cite{Vallilo:2011fj} attempted to use canonical quantization around a massive string solution to calculate the anomalous dimension of a member of the Konishi multiplet
at strong coupling. Although the result agrees with both the prediction from integrability and Green-Schwarz formalism, this approach has several issues that make results unreliable \cite{Heinze:2015xxa}.\footnote{The authors would like to thank Martin Heinze for discussions on the subject.}

An alternative and more desirable approach is to use CFT techniques
to study vertex operators and correlation functions since scattering amplitudes are more easily
calculated using this approach. A first step is to identify physical vertex operators. Since the
pure spinor formalism is based on BRST quantization, physical vertex operators should be in the cohomology
of the BRST charge. For massless states, progress has been made in \cite{Mikhailov:2009rx,Mikhailov:2011af,Berkovits:2012ps,Chandia:2013kja}.
For massive states the computation of the cohomology in a covariant way is a daunting task even in flat space \cite{Berkovits:2002qx}.

A simpler requirement for physical vertices is that they should be primary operators of dimension zero for the unintegrated vertices and primaries of dimension $(1,1)$ in the integrated case. Massless unintegrated vertex operators in the pure spinor formalism are local operators with ghost number $(1,1)$ constructed in terms of zero classical conformal dimension fields \cite{Berkovits:2000fe}. So for them to remain primary when quantum corrections are taken into account, their anomalous dimension should vanish. Massless integrated vertices have zero ghost number and classical conformal dimension $(1,1)$. Therefore they will also be primaries when their anomalous dimension vanishes. Operators of higher mass level are constructed using fields with higher classical conformal dimension. For general mass level $n$ (where $n=1$ corresponds to the massless states) the unintegrated vertex operators have classical conformal dimension $(n-1,n-1)$. If such vertex has anomalous dimension
$\gamma$, the condition for it to be primary is $2n-2 +\gamma =0$. The case for integrated vertex operators is similar. For strings in flat space $\gamma$ is always  $\frac{\alpha' k^2}{2}$, which is the anomalous dimension of the plane wave $e^{ik\cdot X}$. This reproduces the usual mass level formula.

This task of computing $\gamma$ can be
made algorithmic in the same spirit as the four dimensional SYM case \cite{Beisert:2003jj,Beisert:2003yb}.
However, here we are interested in finding the subset of operators satisfying the requirements described above. The value of the energy
of the corresponding string state should come as the solution to an algebraic equation obtained from this requirement.  However we do not expect the energy to be simply one
of the parameters in the vertex operator. The proper way to identify the energy is to compute the conserved charge related to it and apply it to the vertex operator.

In this paper we intend to systematize the computation of anomalous
dimensions in the worldsheet by computing all one-loop logarithmic short distance singularities in
the product of operators with at most two derivatives. To find the answer for operators with more derivatives
one simply has to compute the higher order expansion in the momentum of our basic propagator. We used the method
applied by Wegner in \cite{wegner} for the $O(n)$ model, but modified for the background field method.
This was already used with success
in \cite{Candu:2013cga,Cagnazzo:2014yha} for some ${\mathbb Z}_2 $-super-coset sigma models. The pure spinor string
is a ${\mathbb Z}_4$ coset and it has an interacting ghost system. This makes it more difficult to organize
the dilatation operator in a concise expression and to find a solution to
\begin{equation}\label{first}{\mathfrak D}\cdot {\mathcal O}=0.\end{equation}

We can select a set of ``letters'' $\{\phi^P\}$ among the basic fields of the sigma model, {e.g.} the $AdS$ coordinates, ghosts
and derivatives of these fields. Unlike the case of $\mathcal{N}=4$ SYM, the worldsheet derivative is not one of the elements of the set, so fields with a different number of derivatives correspond to different letters.  Then ${\mathfrak D}$ is of the form

\begin{align}
 {\mathfrak D} =\frac{1}{2} {\mathfrak D}^{P\,Q} \frac{\partial^2}{\partial\phi^P\partial\phi^Q}.
\end{align}

Local worldsheet operators are of the form
\begin{align}
 {\mathcal O} = V_{P\,Q\,R\,S\,T \cdots}\phi^P\phi^Q\phi^R\phi^S\phi^T\cdots,
\end{align}
the problem is to find $V_{A\,B\,C\,D\,E \cdots}$ such that $\mathcal O$ satisfy \eqref{first}. Another important difference with the usual case is that the order of letters does not matter, so $\mathcal O$ is not a spin chain.

The problem of finding physical vertices satisfying this condition will be postponed to a
future publication. Here we will compute $\mathfrak D$ and apply it to some local operators
in the sigma model which should have vanishing anomalous dimension. The search for vertex operators in $AdS$ using
this approach was already discussed in \cite{Tseytlin:2003ac} but without the contribution
from the superspace variables. The author used the same ``pairing'' rules computed in \cite{wegner}.

This paper is organized as follows. In section 2 we describe the
method used by Wegner in \cite{wegner} for the simple case of the
principal chiral field. This method consists of solving a
Schwinger-Dyson equation in the background field expansion. In section
3 we explain how to apply these aforementioned method to the pure spinor $AdS$ string case. The main
derivation and results are presented in the Appendix B. Section 4 contains
applications, where we use our results to compute the anomalous dimension of several conserved currents.
Conclusions and further applications are in section 5.

\section{Renormalization of operators in the principal chiral model}

 The purpose of this section is to review the computation of logarithmic divergences of
 operators in principal chiral models using the background field
 method. Although this is standard knowledge, the approach taken here
 is somewhat unorthodox so we include it for the sake of
 completeness. Also, the derivation of the full propagators in the
 case of $AdS_5\times S^5$ is analogous to what is done in this section,
 so we omit their derivations.

Consider a principal model in some group $G$, with corresponding Lie
algebra $\mathfrak g$, in two dimensions. The action is given by

\begin{equation} S= -\frac{1}{2\pi\alpha^2}\int d^2z\, {\rm Tr}\;
\partial g^{-1} \bar\partial g,\end{equation}

where $\alpha$ is the coupling constant and $g \in G$. Using the
left-invariant currents $J=g^{-1}\partial g$ and defining
$\sqrt{\lambda}=1/\alpha^2$ we can also write

\begin{equation} S=\frac{\sqrt{\lambda}}{2\pi}\int d^2z\, {\rm Tr}\;
J\bar J.
\end{equation}

The full one-loop propagator is derived from the Schwinger-Dyson equation
\begin{equation}\label{sdeq} \left< \delta_z S\,  {\mathcal
      O}(y)\right>
=\langle \delta_z {\mathcal O}(y)\rangle,
\end{equation}
where $\delta_z$ is an arbitrary local variation of the fundamental
fields and ${\mathcal O}(y)$ is a local operator. This equation comes
from the functional integral definition of $\langle\cdots \rangle$. In
order to be more explicit, let us consider a parametrization of $g$
in terms of quantum fluctuations and a classical background
$g= g_0e^X$, where $g_0$ is the classical background,
${ X} = X^a {\mathfrak J}_a$ and ${\mathfrak J}_a \in {\mathfrak g}$
are the generators of the algebra. Then a variation of $g$ is given by
$\delta g= g\delta X$, and $\delta X = \delta X^a {\mathfrak J}_a$
where we have the variation of the independent fields $X^a$. Also, the
variation of some general operator $\mathcal O$ is $\delta {\mathcal
  O} =
\frac{\delta{\mathcal O}}{\delta X^a}  \delta X^a$.
Then we can write the Schwinger-Dyson equation as

\begin{equation} \left< \frac{\delta S}{\delta X^a(z)} {\mathcal O}(y)\right> =
\left< \frac{\delta {\mathcal O}(y)}{\delta X^a(z)}\right>, \end{equation}
and now it is clear that this is a consequence of the identity

\begin{equation}  \int [DX] \frac{\delta}{\delta X^a(z)}\left( e^{-S}
{\mathcal O}(y) \right) =0.
\end{equation}

In the case that ${\mathcal O}(y)= X(y)$ we get the Schwinger-Dyson
equation for the propagator

\begin{equation}\label{sdeq2} \left< \frac{\delta S}{\delta X^a(z)}
X^b(y)\right> = \d_a^b\delta^2(y-z).\end{equation}

This is a textbook way to get the equation for the propagator in free
field theories and our goal here is to solve this equation for the
interacting case at one loop order. The perturbative expansion of the action is done using the background field method. A fixed background $g_0$ is chosen and the quantum fluctuation is defined as $g=g_0e^{X}$. The expansion of the current is given by
\beq J = e^{-X}{\boldsymbol J} e^{X} +  e^{-X}\partial e^{X}, \eeq
where ${\boldsymbol J} = g_0^{-1} \partial g_0$ is the background current.
 At one loop order only quadratic terms
in the quantum field expansion contribute and, as usual, linear terms
cancel by the use of the background equation of motion. This means
that we can separate the relevant terms action in two pieces
$S= S_{(0)} + S_{(2)}$. Furthermore, $S_{(2)}$ contains the kinetic
term plus interactions with the background. So we have

\begin{equation} S= S_{(0)} + S_{kin} + S_{int}. \end{equation}

If we insert this into (\ref{sdeq2}) the terms that depend purely on
the background cancel and we are left with

\begin{equation}\label{sdeq3} \left< \frac{\delta S_{kin}}{\delta
      X^a(z)} X^b(y)+ \frac{\delta S_{int}}{\delta X^a(z)}
    X^b(y)\right> = \d_a^b\delta^2(y-z).\end{equation}

 Since $\frac{\delta S_{kin}}{\delta X^a(z)} =
-{\frac{\sqrt{\lambda}}{2\pi}}\partial\bar\partial X^a(z)$  and
$\frac{\delta S_{int}}{\delta X^a(z)}$ is linear in quantum fields we can write

\begin{equation}\label{sdeq4}
-{\frac{\sqrt{\lambda}}{\pi}}\partial_z\bar\partial_{\bar z}\langle
X^a(z) X^b(y) \rangle +   \int d^2w\frac{\delta^2 S_{int}}
{\delta X^c(w)\delta X^d(z)}\langle X^c(w) X^b(y)\rangle \eta^{ad} =
\eta^{ab}\delta^2(y-z).
\end{equation}

Finally, this is the equation that we have to solve. It is an integral
equation for $\langle X^a(z) X^b(y) \rangle = G^{ab}(z,y)$ which is
the one-loop corrected propagator. The interacting part of the  action is
\begin{equation} S_{int} = {\frac{\sqrt{\lambda}}{2\pi}} \int d^2z
  \left[-\half \Tr( [\partial X,X]\bar {\boldsymbol J} )
-\half \Tr([\bar\partial X,X]{\boldsymbol J} )\right],
\end{equation}
where the boldface fields stand for the background fields.

Now  we calculate\footnote{Using the equation of motion for the
background $\partial \bar {\boldsymbol J} +\bar\partial {\boldsymbol J}=0$.}

\begin{equation}\label{deltaact} \frac{\delta^2 S_{int}}
{\delta X^c(w)\delta X^a(z)} ={\frac{\sqrt{\lambda}}{2\pi}}
[\partial_w\delta^2(w-z) \Tr([T_c,T_a]\bar {\boldsymbol J} )+
\bar\partial_w\delta^2(w-z) \Tr([T_c,T_a] {\boldsymbol J} )],
\end{equation}
which is symmetric under exchange  of $(a,z)$ and $(c,w)$, as
expected. We define $f^{ab}_c=f_{cd}^b\eta^{da}$. So we get the
following equation for the propagator

\begin{equation}\label{propeq} \partial_z\bar\partial_z G^{ab}(z,y) =
-{\frac{\pi}{\sqrt{\la}}}\eta^{ab}\delta^2(y-z)
+{\frac{f^a_{ce}}{2}}\left(\partial_z G^{cb}(z,y))\bar {\boldsymbol J} ^e +
\bar\partial_z G^{cb}(z,y) {\boldsymbol J} ^e\right).
\end{equation}

Performing the Fourier transform
\begin{align}
G^{ac}(z,k)=&\int d^2y e^{-ik\cdot(z-y)}G^{ac}(z,y),
\end{align}
we finally get

\begin{align}
\label{solution}
G^{ab}(z,k)=& \frac{\eta^{ab}}{\sqrt{\la}}{\frac{\pi}{|k|^2}} + \frac{\Box}{|k|^2} G^{ab}(z,k) +i \frac{\partial}{k}G^{ab}(z,k) + i\frac{\bar\partial}{k}G^{ab}(z,k)
\nn \\
            & -f^a_{ce}\bar {\boldsymbol J} ^e \left({\frac{i}{2 \bar k}}+{\frac{\partial_z}{2|k|^2}}\right)G^{cb}(z,k)-f^a_{ce} {\boldsymbol J} ^e\left({\frac{i}{k}}+{\frac{\bar\partial_z}{
    2|k|^2}}\right)G^{cb}(z,k) .
\end{align}

The dependence on one of the coordinates remains because the presence of background fields breaks
translation invariance on the worldsheet. We can solve the equation above iteratively in inverse powers of $k$. The
first few contributions are given by

\begin{align}
G^{ab}(z,k) =& \frac{\eta^{ab}}{\sqrt{\la}}\frac{\pi}{|k|^2}  -\frac{i\pi f^a_{ce}}{2\sqrt{\la}|k|^2}\eta^{cb}\left(\frac{\bar {\boldsymbol J} ^e}{\bar k}+\frac{{\boldsymbol J} ^e}{k}\right)\nonumber \\
             &-\frac{\pi}{4\sqrt{\la}}f^a_{ce}f^c_{df}\eta^{db}\frac{1}{|k|^2}\left(\frac{1}{|k|^2}({\boldsymbol J} ^e\bar {\boldsymbol J} ^f+\bar {\boldsymbol J} ^e {\boldsymbol J} ^f)+\frac{1}{\bar k^2}\bar {\boldsymbol J} ^e\bar {\boldsymbol J} ^f+\frac{1}{ k^2}{\boldsymbol J} ^e {\boldsymbol J} ^f\right)\nonumber\\
             &+\frac{\pi}{2\sqrt{\la}}\frac{\eta^{db}f^a_{df}}{ |k|^2}\left(\frac{1}{ \bar k^2}\bar\partial\bar {\boldsymbol J} ^f+\frac{1}{ k^2}\partial {\boldsymbol J} ^f\right)+\ldots
\end{align}

With this solution we can finally do the inverse Fourier transform,
\begin{align}
G^{ac}(z,y)=&\int \frac{d^2k}{4\pi^2} e^{ik\cdot(z-y)}G^{ac}(z,k),
\end{align}
to calculate $G^{ac}(z,y)$. If we are only interested in the divergent
part of the propagator we can already set $z=y$. 
Furthermore,
selecting only the divergent terms in the momentum integrals we get
\begingroup
\allowdisplaybreaks
\begin{align}
 \langle X^a(z) X^c(z)\rangle =& \frac{{\rm I}\pi}{\sqrt{\la}}\eta^{ac},\\
 \langle X^a(z) \partial X^c(z)\rangle =& -\frac{{\rm I}\pi}{2\sqrt{\la}}\eta^{dc}f_{de}^a {\boldsymbol J}^e,\\
 \langle X^a(z) \bar\partial X^c(z)\rangle =& -\frac{{\rm I}\pi}{2\sqrt{\la}}\eta^{dc}f_{de}^a \bar {\boldsymbol J}^e,\\
 \langle X^a(z) \partial\bar\partial X^b(z)\rangle =& \frac{{\rm I}\pi}{4\sqrt{\la}}\eta^{db}f_{df}^cf_{ce}^a \left({\boldsymbol J} ^e\bar {\boldsymbol J} ^f+\bar {\boldsymbol J} ^e {\boldsymbol J} ^f\right),\\
 \langle X^a(z) \partial\partial X^b(z)\rangle =& -\frac{{\rm I}\pi}{2\sqrt{\la}} \eta^{cb}f_{ce}^a\partial {\boldsymbol J} ^e+\frac{{\rm I}\pi}{4\sqrt{\la}}\eta^{db}f^a_{ce}f^c_{df}{\boldsymbol J} ^e{\boldsymbol J} ^f,\\
 \langle X^a(z) \bar\partial\bar\partial X^c(z)\rangle =& -\frac{{\rm I}\pi}{2\sqrt{\la}} \eta^{cb}f_{ce}^a\bar\partial\bar {\boldsymbol J} ^e+\frac{{\rm I}\pi}{4\sqrt{\la}}\eta^{db}f^a_{ce}f^c_{df}\bar {\boldsymbol J} ^e\bar {\boldsymbol J} ^f,
\end{align}
\endgroup
where
\beq{\rm I}= -\frac{1}{2\pi\epsilon}=\lim_{x\to y}\int \frac{d^{2+\epsilon}}{4\pi^2}\frac{\e^{i k(x-y)}}{|k|^2} \eeq in $d=2+\epsilon$ dimensions, using the standard dimensional regularization \cite{wegner}.
Since $\partial \langle X^a \partial X^c\rangle =
\langle \partial X^a \partial X^c\rangle +
\langle X^a \partial^2 X^c \rangle$ we can further compute
\begin{align}
 \langle \partial X^a(z) \partial X^b(z)\rangle =& -\frac{{\rm I}\pi}{4\sqrt{\la}}(f^a_{ce}f^c_{df}\eta^{db} {\boldsymbol J}^e {\boldsymbol J}^f),\\
 \langle \bar\partial X^a(z) \partial X^b(z)\rangle =& -\frac{{\rm I}\pi}{2\sqrt{\la}}\eta^{cb}f^a_{ce}\bar\partial {\boldsymbol J} ^e-\frac{{\rm I}\pi}{ 4\sqrt{\la}}\eta^{db}f^c_{df}f^a_{ce}(\bar {\boldsymbol J}^e {\boldsymbol J}^f+\bar {\boldsymbol J}^f {\boldsymbol J}^e).
\end{align}

From now on $\langle \cdot \rangle$ will mean only the logarithmically divergent part
of the expectation value. A simple way to extract this information is by defining
\begin{equation} \langle {\mathcal O}\rangle =\half \int d^2z d^2y
\frac{\delta^2{\mathcal O}}{\delta X^a(z)\delta X^b(y)}
\langle X^a(z) X^b(y)\rangle,
\end{equation}
for any local operator $\mathcal O$. Furthermore, we define
\begin{equation} \langle {\mathcal O}\,  ,
{\mathcal O}'\rangle = \int d^2zd^2y \frac{\delta{\mathcal O}}{\delta
  X^a(z)} \frac{\delta{\mathcal O}'}{\delta X^b(y)}
\langle X^a(z) X^b(y)\rangle.
\end{equation}
Following \cite{Tseytlin:2003ac} we will call it ``pairing'' rules. For local operators these two definitions always give two delta functions, effectively setting all fields at the same point. So the computation of $\langle \cdot\rangle$ can be summarized as
\beq  \langle {\mathcal O} \rangle = \frac{1}{2} \langle X^a X^b\rangle \frac{\partial^2}{\partial X^a \partial X^b} {\mathcal O} = {\mathfrak D} {\mathcal O},\eeq
where
\beq {\mathfrak D} = \frac{1}{2} \langle X^a X^b\rangle \frac{\partial^2}{\partial X^a \partial X^b} \eeq
is the dilatation operator. We can also define $\langle \cdot\, ,\cdot \rangle$ as
\beq  \langle {\mathcal O}\, , {\mathcal O}'\rangle =  \langle X^a X^b\rangle
\frac{\partial {\mathcal O}}{\partial X^a}\frac{\partial {\mathcal O}'}{\partial X^b}.
\eeq

With the above definitions, the divergent part of any product of local operators at the same point
can be computed using.
\begin{equation} \langle {\mathcal O} {\mathcal O}' \rangle=
\langle {\mathcal O}\rangle{\mathcal O}'+ {\mathcal O}
\langle {\mathcal O}' \rangle + \langle {\mathcal O}\,
, {\mathcal O}'\rangle.
\end{equation}
Several known results can be derived using this simple set of rules. Following this procedure in the case of the symmetric space $SO(N+1)/SO(N)$ gives the same results obtained by Wegner \cite{wegner} using a different method.

\section{Dilatation operator for the $AdS_5\times S^5$ superstring}

In this section we will apply the same technique to the case of the pure
spinor $AdS$ string. We begin with a review of the pure spinor
description, pointing out the differences between this model and the principal chiral model, and then describe the main steps of the computation.

\subsection{Pure spinor $AdS$ string}

The pure spinor string \cite{Berkovits:2002qx,Vallilo:2003nx,Berkovits:2004xu} in $AdS$ has the
same starting point as the Metsaev-Tseyltin \cite{Metsaev:1998it}. The maximally supersymmetric type IIB background $AdS_5\times S^5$ is described by the supercoset
\begin{align}
\frac{G}{H}=\frac{PSU(2,2|4)}{SO(1,4)\times SO(5)}.
\end{align}

The pure spinor action is given by
\begin{align}
 S_{PS}=&\frac{R^2}{2\pi}\int d^2z \text{STr}\left[\frac{1}{2}J_2\bar J_2+\frac{1}{4}J_1\bar J_3+\frac{3}{4}\bar J_1 J_3+\o \bar\P\l+\ho\P\hl-N\hat N\right], \label{PSa}
\end{align}
where
\begin{align}
 \P \cdot=&\p \cdot+[J_0,\cdot],&
 N=&\{\o,\l\},&
 \hat N=&\{\ho,\hl\}.
\end{align}


There are several difference between the principal chiral model action and (\ref{PSa}). First, the model is coupled to ghosts. The pure spinor action also contains a Wess-Zumino term, and the global invariant current $J$ belongs to the $\mathfrak{psu}(2,2|4)$ algebra, which is a graded algebra, with grading 4. Thus we split the current as $J=A+J_1+J_2+J_3$, where $A=J_0$ belongs to the algebra of the quotient group $H=SO(1,4)\times SO(5)$. The notation that we use for currents of different grade is
\begin{align}
 J_0=& J_0^i T_i \quad;\quad J_1=J_1^\a T_\a \quad;\quad J_2=J_2^m T_m \quad;\quad J_3=J_3^\ha T_\ha.
\end{align}

The ghosts fields are defined as
\begin{align}
\l=&\l^A T_A\quad;\quad \o=-\o_A\eta^{A\hA}T_\hA\quad;\quad \hl=\hl^\hA T_\hA \quad;\quad \ho=\ho_\hB\eta^{B\hB}T_B.
\end{align}
Note that $A$ and $A'$ indices on the ghosts mean $\a$ and $\ha$, but we will use a different letter in order to make it easier to distinguish
which terms come from ghosts and which come from the algebra. The pure spinor condition can be written as
\beq\label{pureeqs} \{\l, \l\} = \{\hat\l,\hat\l\} =0.\eeq

Following the principal chiral model example, we expand $g$ around a classical background $g_0$ using the $g=g_0 e^X$ parametrization. It is worth noting that $X=x_0+x_1+x_2+x_3$ belongs to the $\mathfrak{psu}(2,2|4)$ algebra, but we can use the coset property to fix $x_0=0$. With this information the quantum expansion of the left invariant current is
\begin{align}
\begin{split}
 A=&{\boldsymbol A} + \sum_{i=1}^3\left(\left[\boldsymbol{J}_i,x_{4-i}\right]+\frac{1}{2}\left[\nabla x_i,x_{4-i} \right]\right)+\sum_{i,j=1}^{3}\left[ \left[\boldsymbol{J}_i,x_j \right],x_{8-i-j}\right],\\
 J_l=&\boldsymbol{J}_l +\nabla x_l + \sum_{i=1}^3 \left(\left[\boldsymbol{J}_i,x_{4+l-i}\right]+\frac{1}{2}\left[\nabla x_i,x_{4+l-i} \right]\right)+\sum_{i,j=1}^{3}\left[ \left[\boldsymbol{J}_i,x_j \right],x_{8+l-i-j}\right],\\
 \l=&\boldsymbol{\l}+\d\l,\\
 \o=&\boldsymbol{\o}+\d\o,\\
 \hat\l=&\boldsymbol{\hat\l}+\d\hat\l,\\
 \hat\o=&\boldsymbol{\hat\o}+\d\hat\o.
\end{split} \label{jexp}
\end{align}

Where we take $x_0=0$ as mentioned before, and we used $g_0^{-1}\partial g_0=\boldsymbol{J}=\boldsymbol{A}+\boldsymbol{J}_1+\boldsymbol{J}_2+\boldsymbol{J}_3$. The boldface terms stand for the background term, both for the currents and for the ghost fields. 

Using all this information inside the action we get
\begin{align}
 S_{PS}=\frac{R^2}{2\pi}\int d^2z \left[\frac{1}{2}\P x_2^m \bar\P x_2^n \eta_{mn}-\P x_1^\a \bar \P x_3^\ha \eta_{\a\ha}+\d\o_A\bar\p\d\l^A+\d\ho_\hA \p\d\hl^\hA\right]+S_{int}. \label{exp1}
\end{align}

The full expansion can be found in the Appendix C. In order to compute the logarithmic  divergences, we need to generalize the method explained in section 2 for a coset model with ghosts. The following subsection is devoted to explain this generalization.

\subsection{General coset model coupled to ghosts}

 In this subsection we generalize the method of Section 2 to the case of a general coset $G/H$ and then specialize for the pure spinor string case. We will denote the corresponding algebras $\mathfrak g$ and $\mathfrak h$, where $\mathfrak h$ should be a subalgebra of $\mathfrak g$. The generators of ${\mathfrak g}-{\mathfrak h}$ will be denoted by $T_a$ where $a=1$ to ${\rm dim}_{\mathfrak g} -{\rm dim}_{\mathfrak h}$ and the generators of $\mathfrak h$ will be denoted by $T_i$ where $i=1$ to ${\rm dim}_{\mathfrak h}$. We also include a pair of first order systems $(\lambda^A,\omega_B)$  and $(\hat\lambda^{A'},\hat\omega_{B'})$ transforming in two representations $(\Gamma^i{}^B_A,\Gamma^i{}^{B'}_{A'})$ of $\mathfrak h$. We will assume that the algebra $\mathfrak g$ has the following commutation relations
 \beq\label{cosetalg} [ T_a,T_b]= f^c_{ab} T_c + f^i_{ab} T_i, \quad [T_a,T_i]= f^b_{ai}T_b,\quad [T_i,T_j]=f^k_{ij} T_k,\eeq
where $f^c_{ab}\neq 0$ for a general coset and $f^c_{ab}=0$ if there is a $\mathbb Z_2$ symmetry, i.e., $G/H$ is a symmetric space.
As in the usual sigma model $g \in G/H$ and the currents $J=g^{-1}\partial g$ are invariant by left global transformations in $G$. We can decompose
$J= J^aT_a+ A^iT_i$ where $ J^a T_a\in{\mathfrak g}-{\mathfrak h}$ and $ A^i T_i\in{\mathfrak h}$. With this decomposition $K$ transforms in the adjoint representation of $\mathfrak h$ and $A$ transforms as a connection. We will also allow a quartic interaction in the first order sector. Defining $N^i= \lambda^A \Gamma^i{}^B_A\omega_B$ and $\hat N^i=\hat\lambda^{A'} \Gamma^i{}^{B'}_{A'}\hat\omega_{B'}$, the interaction will be $\beta  N^i \hat N_i$ where $\beta$ is a new coupling constant that in principle is not related with the sigma model coupling.

The total action is given by
\beq S= \int d^2z \left( \,\Tr\; (J-A)(\bar J-\bar A) + \omega_A \bar\nabla \lambda^A + \hat\omega_{A'}\nabla\lambda^{A'} + \beta N^i\hat N_i \right),\eeq
where $(\nabla,\bar\nabla)=(\partial - A^i\Gamma_i{}^{A'}_{B'}\, , \bar\partial - \bar A^i\Gamma_i{}^A_B)$ are the covariant derivatives for the first order system ensuring gauge invariance.

The background field expansion is different if we are in a general coset or a symmetric space. Since we want to generalize the results to the case of $AdS_5\times S^5$, we will use a notation that keeps both types of interactions. Again, the quantum coset element is written as $g=g_0 e^X$ where $g_0$ is the classical background and $X= X^aT_a$ are the quantum fluctuations. Up to quadratic terms in the quantum fluctuation the expansion of the action is
\begin{align}
S =& S_0 + \int d^2z \left(  \eta_{ab}\nabla X^a \bar\nabla X^b -  Z_{abc} \boldsymbol{J}^a X^b \bar\nabla X^c
-\bar Z_{abc} \bar{\boldsymbol{J}}^a X^b \nabla X^c + R_{abcd} \boldsymbol{J}^a\bar {\boldsymbol{J}}^b X^cX^d\right. \nonumber\\
   &\left. + \delta\omega_A \bar \partial \delta\lambda^A + \delta\hat\omega_{A'}\partial\delta\hat\lambda^{A'} + \bar{\boldsymbol A}^i N_i + {\boldsymbol A}^i\hat N_i+ \beta \left(\{\delta \l,\boldsymbol{\o}\}+\{\boldsymbol{\l},\delta\o\} \right)^i \left(\{\delta \hat\l,\hat{\boldsymbol{\o}}\}+\{\hat{\boldsymbol{\l}},\delta\hat\o\} \right)_i \right),
\end{align}
where the covariant derivatives on $X$ are $(\partial -[A,\; \cdot\; ],\bar\partial - [\bar A, \;\cdot\; ])$. The tensors $(Z_{abc},\bar Z_{abc}, R_{abcd})$ appearing above are model dependent. In the case of a symmetric space $Z=\bar Z=0$ and $R_{abcd} = f^i_{ab}f_{icd}$. In the general coset case $Z_{abc}=\bar Z_{abc}= \half f_{abc}$. If there is a Wess-Zumino term, the values of $Z_{abc}$ and $\bar Z_{abc}$ can differ.  Since we want to do the general case, we will not substitute the values of these tensor until the end of the computations. In the action above the quantum connections
have the following expansion
\beq { A}^i =\boldsymbol{A}^i +f^i_{ab}\boldsymbol{J}^aX^b + \half f^i_{ab} \nabla X^a X^b +W^i_{abc}\boldsymbol{J}^aX^bX^c +\cdots  \eeq
\beq \bar{A}^i =\bar{\boldsymbol{A}}^i +f^i_{ab}\bar{\boldsymbol{J}}^aX^b + \half f^i_{ab} \bar\nabla X^a X^b +W^i_{abc}\bar{\boldsymbol{J}}^aX^bX^c +\cdots  \eeq

where $W^i_{abc}=\half f^d_{ab}f^i_{dc}$ for a general coset and vanishes for a symmetric space.

To proceed, we have to compute the second order variation of the action with respect to the quantum fields. The difference this time is that there are many more couplings, so we expect a system of coupled Schwinger-Dyson equations, corresponding to each possible corrected propagator. For example, in the free theory approximation there is no propagator between the sigma model fluctuation and the first order system, but due to the interactions there we may have corrected propagators between them.


Since a propagator is not a gauge invariant quantity, it can depend on gauge dependent combinations of the background gauge fields $(A^i, \bar A^i)$. Furthermore, since we have chiral fields transforming in two different representations of $\mathfrak h$ it is possible that the quantum theory has anomalies. In the case of the $AdS_5\times S^5$ string sigma-model it was argued by Berkovits \cite{Berkovits:2004xu} that there is no anomaly for all loops.  An explicit one loop computation was done in \cite{Mazzucato:2009fv}. Therefore it is safe to assume that the background gauge fields only appear in physical quantities in a gauge invariant combination. The simplest combination of this type is $\Tr [\nabla,\bar\nabla]^2$. Since the classical conformal dimension of this combination is four and so far we are interested in operators of classical conformal dimension 0 and 2, we can safely ignore all interactions with $(A^i,\bar A^i)$.

We will assume a linear quantum variation of the first order system, e.g., $\lambda^A \to \boldsymbol{\lambda}^A + \delta\lambda^A$. Instead of introducing more notation and a cumbersome interaction Lagrangian, we will simply compute the variations of these fields in the action and set to their background values the remaining fields.

With all these simplifications and constraints in mind, let us start constructing the Schwinger-Dyson equations. First we compute all possible non-vanishing second variations of the action
\begingroup
\allowdisplaybreaks
\begin{subequations} \label{dd}
\begin{align}
 \frac{\delta^2S_{int}}{\delta X^a\delta X^b} &=\delta^2(z-w)\left[ {\boldsymbol J}^ c\bar {\boldsymbol J}^ d( R_{cdab} + R_{cdba}) + {\boldsymbol N}_ i\bar {\boldsymbol J}^ c( W^i_{cab}+W^i_{cba}) + \hat {\boldsymbol N}_ i {\boldsymbol J}^ c( W^i_{cab}+W^i_{cba})\right] \nonumber \\
                                               &-\partial_w\delta^2(z-w)[ \bar Z_{cab} \bar {\boldsymbol J}^ c + f^i_{ab} \hat {\boldsymbol N}_ i] -\bar\partial_w \delta^2(z-w) [ Z_{cab} {\boldsymbol J}^ c + f^i_{ab} {\boldsymbol N}_ i],\\
 \frac{\delta^2S_{int}}{\delta\lambda^A\delta\omega_B} &=\delta^2(z-w) \beta \Gamma^i{}^B_A \hat {\boldsymbol N}_ i, \\
 \frac{\delta^2S_{int}}{\delta\lambda^A\delta X^a} &= \delta^2(z-w)(\Gamma_i{\boldsymbol \omega})_A f^i_{ba}\bar {\boldsymbol J}^ b, \\
 \frac{\delta^2S_{int}}{\delta\lambda^A\delta\hat\lambda^{B'}} &= \delta^2(z-w)\beta (\Gamma^i{\boldsymbol \omega})_A(\hat\Gamma_i\hat{\boldsymbol \omega})_{B'}, \\
 \frac{\delta^2S_{int}}{\delta\lambda^A\delta\hat\omega_{B'}} &= \delta^2(z-w)\beta (\Gamma^i{\boldsymbol \omega})_A(\hat{\boldsymbol \lambda}\hat\Gamma_i)^{B'}, \\
 \frac{\delta^2S_{int}}{\delta\hat\lambda^{A'}\delta\hat\omega_{B'}} &= \delta^2(z-w)\beta {\boldsymbol N}^i \hat\Gamma_i{}^{B'}_{A'}, \\
 \frac{\delta^2S_{int}}{\delta\hat\lambda^{A'}\delta X^a} &=\delta^2(z-w) (\hat\Gamma_i\hat{\boldsymbol\omega})_{A'}f^i_{ba}{\boldsymbol J}^ b, \\
 \frac{\delta^2S_{int}}{\delta\hat\lambda^{A'}\delta\omega_{B}} &= \delta^2(z-w)\beta ({\boldsymbol\lambda}\Gamma^i)^B(\hat\Gamma_i\hat{\boldsymbol\omega})_{A'}, \\
 \frac{\delta^2S_{int}}{\delta X^a\delta\omega_{B}} &= \delta^2(z-w)({\boldsymbol\lambda}\Gamma_i)^B f^i_{ba} \bar {\boldsymbol J}^ b, \\
 \frac{\delta^2S_{int}}{\delta X^a\delta\hat\omega_{B'}} &=\delta^2(z-w) (\hat{\boldsymbol\lambda}\hat\Gamma_i)^{B'} f^i_{ba} {\boldsymbol J}^ b,  \\
 \frac{\delta^2S_{int}}{\delta\omega_A\delta\hat\omega_{B'}} &= \delta^2(z-w)\beta ({\boldsymbol\lambda}\Gamma^i)^A(\hat{\boldsymbol\lambda}\hat\Gamma_i)^{B'}.
\end{align}
\end{subequations}
\endgroup

We are going to denote these second order derivatives generically as $I_{\Sigma\Lambda}(z,w)$ where $\Sigma$ and $\Lambda$ can be any of the indices $({}^a,{}^A,{}_B,{}^{A'},{}_{B'})$. Also, the quantum fields will be denoted by $\Phi^\Sigma$(z). With this notation the Schwinger-Dyson equations are
\beq\label{gensde} \langle\frac{\delta S_{kin}}{\delta \Phi^\Lambda(z)} \Phi^\Sigma(y) \rangle + \int d^2w \frac{\delta^2 S_{int}}{\delta\Phi^\Upsilon(w)\delta\Phi^\Lambda(z)}  \langle \Phi^\Upsilon(w) \Phi^\Sigma(y)\rangle =\delta^\Sigma_\Lambda \delta(z-y). \eeq

Note that the only non-vanishing components of $\delta^\Sigma_\Lambda$ are $\eta^{ab}$, $\delta^A_B$ and $\delta^{A'}_{B'}$. Since the type and the position of the indices completely identity the field, the propagators are going to be denoted by $G^{\Sigma\Lambda}(z,y)=   \langle \Phi^\Sigma(z)\Phi^\Lambda(y)\rangle $.
Since we five different types of fields, we have fifteen coupled Schwinger-Dyson equations to solve. Again we have to make a simplification. Interpreting $(\lambda^A,\hat\lambda^{A'})$ as left and right moving ghosts and knowing that in the pure spinor superstring unintegrated vertex operators have ghost number $(1,1)$
with respect to $(G,\hat G)$, we will concentrate on only four corrected propagators $\langle X^a(z) X^b(y)\rangle$, $\langle X^a(z)\lambda^A(y)\rangle$, $\langle X^a(z)\hat\lambda^{A'}(y)\rangle$ and $\langle\lambda^A(z)\hat\lambda^{A'}(y)\rangle$. As in the principal chiral model case we are going to solve the Schwinger-Dyson equations first in momentum space. It is useful to note that since we will solve this equations in inverse powers of $k$, the first contributions to the corrected propagators will have the form
\beq \langle X^c X^a \rangle \approx \frac{\eta^{ca}}{|k|^2}\, , \qquad \langle \omega_A\lambda^B\rangle \approx \frac{\delta_A^B}{\bar k}\, ,\qquad \langle \hat\omega_{A'}\hat\lambda^{B'}\rangle \approx \frac{\delta_{A'}^{B'}}{ k}.\eeq

Regarding $(A,A')$ as one type of index we can arrange the whole Schwinger-Dyson equation into a matrix notation with three main blocks. Doing the same Fourier transform as before we get a matrix equation that can be solved iteratively
\beq G_\Sigma^\Upsilon = I^\Upsilon_\Sigma + ( F_{\Sigma\Gamma} + \Delta_{\Sigma\Gamma}) G^{\Gamma\Upsilon} \label{green},\eeq
where
\beq I^\Upsilon_\Sigma = \left(\begin{array}{ccc} \frac{\delta^a_b}{|k|^2} & 0 & 0\\
0&0& -i\frac{ \delta^A_B}{k}\\ 0& i\frac{\delta^B_A}{k}&0  \end{array}\right),
\eeq
\beq F_{\Sigma\Gamma} = \frac{\delta^2 S_{int}}{\delta\Phi^\Sigma\delta\Phi^\Gamma},\qquad \Delta_{\Sigma\Gamma} =\left(\begin{array}{ccc} \frac{\partial\bar\partial}{|k|^2} +i\frac{\partial}{k}+i\frac{\bar\partial}{\bar k} &0&0\\
0& -i\frac{\partial}{k} & 0\\ 0&0& i\frac{\partial}{k}\end{array}\right) .
\eeq
All elements of the interaction matrix $F_{\Sigma\Gamma}$ are shown in Appendix C.
As in section 2, the solution to equation (\ref{green}) is computed iteratively
\beq G^{(0)}{}^\Upsilon_\Sigma =  I^\Upsilon_\Sigma,\qquad G^{(1)}{}^\Upsilon_\Sigma = F_{\Sigma}^{\Gamma}I_\Gamma^{\Upsilon},\eeq
and so on for higher inverse powers of $k$.

\subsection{Pairing rules}

As discussed in the introduction and Section 2, the computation of the divergent part of any local operator
can be summarized by the pairing rules of a set of letters $\{\phi^P\}$. The complete set of these pairing rules can be found in
the Appendix C. If we choose a set of letters such that $\langle \phi^P \rangle =0$, then the divergent part of the product
of two letters is simply
\begin{align}
\langle \phi^P \phi^Q \rangle =\langle \phi^P,\; \phi^Q \rangle.
\end{align}

We computed the momentum space Green function up to quartic inverse power of momentum so we must restrict our set of
letters to fundamental fields up to classical dimension one. The convenient set of letter we will use is
\begin{align}
\{\phi^P\} =\{x_2^a,x_1^\alpha,x_3^{\hat\alpha},J_2^a,J_1^\alpha,J_3^{\hat\alpha},J_0^{i},\bar J_2^a,\bar J_1^\alpha,\bar J_3^{\hat\alpha},\bar J_0^{i},\lambda^A,\omega_A,\hat\lambda^{\hat A},\hat\omega_{\hat A}, N^i,\hat N^i\}.
\end{align}
If we extend the computation to take into account operators with more than two derivatives the set of letters has to be extended to include them. The matrix elements of the dilation
operator ${\mathfrak D}^{P\,Q} = \langle \phi^P, \phi^Q\rangle$ are the full set of pairings described in Appendix C.4. To avoid cumbersome notation, the pairing rules are written contracting with the corresponding $\mathfrak{psu}(2,2|4)$ generator. The computations done in next section are a straightforward application of the differential operator
\begin{align}\label{DilatationD}
 {\mathfrak D} =\frac{1}{2} {\mathfrak D}^{P\,Q} \frac{\partial^2}{\partial\phi^P\partial\phi^Q}
\end{align}
on a a local operator of the form ${\mathcal O} = V_{P\, Q\, R\, S\, T \cdots}\phi^P\phi^Q\phi^R\phi^S\phi^T\cdots$.

\section{Applications}

In this section we use our results to prove that certain important operators in the pure spinor sigma model are not renormalized. The operators we choose are stress energy tensor, the conserved currents related to the global $PSU(2,2|4)$ symmetry
and the composite $b$-ghost. All these operators are a fundamental part of the formalism and it is a consistency check that they are indeed not renormalized. All the computations bellow are an application of the differential operator (\ref{DilatationD}).  We use the notation $\langle {\cal O}\rangle = {\mathfrak D}\cdot {\cal O}$.

\subsection{Stress-energy tensor}

The holomorphic and anti-holomorphic stress-energy tensor for (\ref{PSa}) are given by

\begin{align}
 T=&\text{STr}\left(\frac{1}{2}J_2J_2+J_1J_3+\o\nabla\l\right),\\
 \bar T=&\text{STr}\left(\frac{1}{2}\bar J_2 \bar J_2+\bar J_1\bar J_3+\ho\bar \nabla\hl\right).
\end{align}

For the holomorphic one
\begingroup
\allowdisplaybreaks
\begin{align}
 \langle T \rangle =& \text{STr}\left(\frac{1}{2}\langle J_2,J_2 \rangle+\langle J_1,J_3 \rangle - N\langle J_0 \rangle\right)\nonumber\\
                   =&\text{STr}\left(\frac{1}{2} [{\boldsymbol N},T_m][{\boldsymbol N},T_n]\eta^{mn}-[{\boldsymbol N},T_\a][{\boldsymbol N},T_\ha]\eta^{\a\ha}\right.\nonumber\\&\left.+\frac{1}{2}{\boldsymbol N}\left(\{[{\boldsymbol N},T_\ha],T_\a \}\eta^{\a\ha}-\{[{\boldsymbol N},T_\a],T_\ha\}\eta^{\a\ha}+[[{\boldsymbol N},T_m],T_n]\eta^{mn}\right)\right)\nonumber\\
                   =&0.
\end{align}
\endgroup
We used the results in (\ref{<j0>},\ref{<j1j3>}) and the identity (\ref{a}). A similar computations happens to the antiholomorphic $\bar T$, where now we use the results in (\ref{<j0b>},\ref{<j1bj3b>}) and the identity (\ref{b}).

\subsection{Conserved currents}

The string sigma model is invariant under global left-multiplications by an element of
$\mathfrak{psu}(2,2|4)$, $\delta g = \Lambda g$. We can calculate the conserved currents related to this symmetry using standard Noether method. The currents are given by
\begin{align}
 j=&g\left(J_2+\frac{3}{2}J_3+\frac{1}{2}J_1-2N\right)g^{-1}=g A g^{-1},\\
 \bar j=&g\left(\bar J_2+\frac{1}{2}\bar J_3+\frac{3}{2}\bar J_1-2\hat N\right)g^{-1}=g \bar A g^{-1}.
\end{align}

They should be free of divergences. To see that this is the case, it is easier to compute by parts:
\begin{align}
 \langle j \rangle =&\langle g \rangle {\boldsymbol A} g_0^{-1} + \langle g,A \rangle  g_0^{-1}+\langle g {\boldsymbol A}, g^{-1}\rangle+  g_0 \langle A, g^{-1}\rangle +g_0 {\boldsymbol A} \langle g^{-1}\rangle + g_0 \langle A\rangle g_0^{-1}.
\end{align}

We have defined $\langle A {\boldsymbol B}, C \rangle$ as usual, but taking ${\boldsymbol B}$ as a classical field, thus $\langle A {\boldsymbol B}, C \rangle=\langle A, {\boldsymbol B} C\rangle$. From (\ref{j}) we get $\langle A\rangle=0$, and using (\ref{xx}) we obtain
\begin{align}
 \langle g \rangle {\boldsymbol A} g_0^{-1}+\langle g {\boldsymbol A}, g^{-1} \rangle+ g_0 {\boldsymbol A}\langle g^{-1} \rangle=&\frac{1}{2}g_0 \langle \left[\left[{\boldsymbol A}, X\right],X\right] \rangle g_0^{-1}\nonumber\\
 =&\frac{1}{2}g_0\left([\left[{\boldsymbol A},T_m \right],T_n]\eta^{mn}+\{[{\boldsymbol A},T_\ha],T_\a\}\eta^{\a\ha}\right.\nonumber\\
  &\left.-\{[{\boldsymbol A},T_\a],T_\ha\}\eta^{\a\ha}\right)g_0^{-1}.
\end{align}

For the currents, using the results (\ref{xj1}-\ref{xj3b}),
\begin{align}
 g_0^{-1}\langle g, J_1 \rangle + \langle J_1,g^{-1}\rangle g_0=&-\{[{\boldsymbol J}_2,T_\ha],T_\a\}\eta^{\a\ha}-\{[{\boldsymbol J}_3,T_\ha],T_\a\}\eta^{\a\ha}+\{[{\boldsymbol N},T_\ha],T_\a\}\eta^{\a\ha},\\
 g_0^{-1}\langle g, J_2 \rangle + \langle J_2,g^{-1}\rangle g_0=& -[[{\boldsymbol J}_3,T_m],T_n]\eta^{mn}+[[{\boldsymbol N},T_m],T_n]\eta^{mn},\\
 g_0^{-1}\langle g, J_3 \rangle + \langle J_3,g^{-1}\rangle g_0=& -\{[{\boldsymbol N},T_\a],T_\ha\}\eta^{\a\ha},\\
 g_0^{-1}\langle g, N \rangle + \langle N,g^{-1}\rangle g_0=& 0,
\end{align}
but we already know that $\left\{ \left[ J_{1,3}, T_a\right], T_b\right\}g^{ab}=0$, for $a=\{i,m,\a,\ha\}$, see (\ref{c}). Thus,
\begin{align}
 g_0^{-1}\langle j\rangle g_0=& - \frac{1}{2}\left( \left\{\left[{\boldsymbol N},T_\ha \right],T_\a\right\} + \left\{\left[{\boldsymbol N},T_\a \right],T_\ha\right\} \right)\eta^{\a\ha}\nonumber\\
                              &+\frac{1}{2} \left( \left\{\left[{\boldsymbol J}_2,T_m \right],T_n\right\}\eta^{mn} - \left\{\left[{\boldsymbol J}_2,T_\a \right],T_\ha\right\} \eta^{\a\ha} \right).
\end{align}

By lowering all the terms in the structure coefficients, we can see that the first term is just $(f_{i\a\hb}f_{j\ha\b}-f_{i\ha\b}f_{j\a\hb})\eta^{\a\ha}\eta^{\b\hb}$, and the second term is proportional to the dual coxeter number, see (\ref{a},\ref{b}), which is 0. Thus, summing everything, we get
\begin{align}
 \langle j\rangle =0.
\end{align}

For the antiholomorphic current we just obtain, using the same results as before,
\begin{align}
 g_0^{-1}\langle g,\bar J_1 \rangle + \langle \bar J_1,g^{-1}\rangle g_0=& \{[\hat {\boldsymbol N},T_\ha],T_\a\}\eta^{\a\ha},\\
 g_0^{-1}\langle g,\bar J_2 \rangle + \langle\bar J_2,g^{-1}\rangle g_0=& -[[\bar {\boldsymbol J}_1,T_m],T_n]\eta^{mn}+[[\hat {\boldsymbol N},T_m],T_n]\eta^{mn},\\
 g_0^{-1}\langle g,\bar J_3 \rangle + \langle\bar J_3,g^{-1}\rangle g_0=& \{[\bar {\boldsymbol J}_1,T_\a],T_\ha\}\eta^{\a\ha}+\{[\bar {\boldsymbol J}_2,T_\a],T_\ha\}\eta^{\a\ha}-\{[\hat {\boldsymbol N},T_\a],T_\ha\}\eta^{\a\ha},
\end{align}
and using $\left\{ \left[ J_{1,3}, T_a\right], T_b\right\}g^{ab}=0$ we see that doing the same as $j$, we arrive at $\langle \bar j\rangle=0$.

\subsection{$b$ ghost}

The pure spinor formalism does not have fundamental conformal ghosts. However, in a consistent string theory, the stress-energy tensor must be BRST exact $T=\{Q,b\}$. So there must exist a composite operator of ghost number $-1$ and conformal weight $2$. The flat space $b$-ghost was first computed in \cite{Berkovits:2004px} and a simplified expression for it in the $AdS_5\times S^5$ background was derived in \cite{Berkovits:2008ga}. In our notation, the left and right moving $b$-ghosts can be written as
\begin{align}
 b=&(\l\hl)^{-1}\text{STr}\left(\hl [J_2,J_3]+\{\o,\hl\}[\l,J_1]\right)-\text{STr}\left(\o J_1\right),\\
 \bar b=&(\l\hl)^{-1}\text{STr}\left(\l [\bar J_2,\bar J_1]+\{\ho,\l\}[\hl,\bar J_3]\right)-\text{STr}\left(\ho \bar J_3\right),
\end{align}
where $(\l\hl)=\l^A \hl^\hA \eta_{A\hA}$.

Let us first compute the divergent part of the left moving ghost; we will need the results from (\ref{ghost1}) to (\ref{ghost2}):
\begin{align}
 \langle b \rangle =& ({\boldsymbol\l}\hat {\boldsymbol\l})^{-1}\text{STr}\langle\hl [J_2,J_3]+\{\o,\hl\}[\l,J_1]\rangle-(\l\hl)^{-2}\langle \l\hl \rangle \text{STr}\left(\hl [J_2,J_3]+\{\o,\hl\}[\l,J_1]\right)\nonumber\\& -({\boldsymbol\l}\hat{\boldsymbol\l})^{-2}\langle (\l\hl),\text{STr}\left(\hl [J_2,J_3]+\{\o,\hl\}[\l,J_1]\right)\rangle-\text{STr}\langle\o J_1\rangle,
\end{align}

The $\langle \l\hl\rangle$ term is easy,
\begin{align}
 \langle (\l\hl)\rangle=&-{\boldsymbol\l}^A \hat{\boldsymbol\l}^\hA f^B_{Ai} f^\hB_{\hA j}g^{ij}\eta_{B\hB}=0;
\end{align}
where we have used (\ref{c}).
The $\langle \o J_1\rangle$ term is also 0. The other terms are
\begin{align}
 \text{STr}\langle\hl [J_2,J_3] \rangle=&-\text{STr}\left([\hat {\boldsymbol\l},T_i] ([[{\boldsymbol J}_2,T_j],{\boldsymbol J}_3]+[{\boldsymbol J}_2,[{\boldsymbol J}_3,T_j]])g^{ij} \right)\nonumber\\
                                       =&-\text{STr}\left([\hat {\boldsymbol\l},T_i][T_j,[{\boldsymbol J}_2,{\boldsymbol J}_3]]g^{ij}\right)=-\text{STr}\left([\left[\hat {\boldsymbol\l},T_i \right],T_j][{\boldsymbol J}_2,{\boldsymbol J}_3]g^{ij}\right)=0 ,
\end{align}
we used $f_{i\a\hb}f_{j\ha\b}g^{ij}\eta^{\a\ha}=0$, see (\ref{c}).
The next term is
\begin{align}
 \text{STr}\langle \{\o,\hl\}[\l,J_1] \rangle=& -\text{STr}\left(\{[{\boldsymbol\o},T_i],[\hat {\boldsymbol\l},T_j]\}[{\boldsymbol\l},{\boldsymbol J}_1]+\{{\boldsymbol \o},[\hat {\boldsymbol\l},T_i]\}[\left[{\boldsymbol\l},T_j\right],{\boldsymbol J}_1]\right.\nonumber\\
                                              &\left. +\{{\boldsymbol\o},[\hat {\boldsymbol\l},T_i]\}[{\boldsymbol\l},[{\boldsymbol J}_1,T_j]] \right)g^{ij}\nonumber\\
                                             =&-\text{STr}\left(\{{\boldsymbol\o},[\hat {\boldsymbol\l},T_i]\}([T_j,[\l,{\boldsymbol J}_1]]+[[{\boldsymbol \l},T_j],{\boldsymbol J}_1] +[{\boldsymbol \l},[{\boldsymbol J}_1,T_j]]) \right)g^{ij}\nonumber\\
                                             =&0,
\end{align}
which comes from the Jacobi identity, see appendix B. The remaining terms are computed using
\begin{align}
 \l^A [\hl^\hA,T_i]\eta_{A\hA}=&-[\l^A,T_i]\hl^\hA \eta_{A\hA}=\{\l,\hl\}^j g_{ij},
\end{align}
thus
\begin{align}
 \langle (\l\hl),\text{STr}\left(\hl [J_2,J_3]\right)\rangle=&\text{STr}\left([\hat{\boldsymbol\l},\{{\boldsymbol\l},\hat{\boldsymbol\l}\}] [{\boldsymbol J}_2,{\boldsymbol J}_3]\right)+\text{STr}\left(\hat {\boldsymbol\l} [[{\boldsymbol J}_2,{\boldsymbol J}_3],\{{\boldsymbol\l},\hat {\boldsymbol\l}\}]\right)=0,\\
 \langle (\l\hl),\text{STr}\left(\{\o,\hl\}[\l,J_1]\right)\rangle=&\text{STr}\left(\{{\boldsymbol\o},[\hat {\boldsymbol\l},\{{\boldsymbol\l},\hat {\boldsymbol\l}\}]\}[{\boldsymbol\l},{\boldsymbol J}_1]-\{[{\boldsymbol \o},\{{\boldsymbol \l},\hat {\boldsymbol\l}\}],\hat {\boldsymbol\l}\}[{\boldsymbol\l},{\boldsymbol J}_1]\right.\nonumber\\
                                                                  &\left.+[\{{\boldsymbol \o},\hat{\boldsymbol \l}\},\{ {\boldsymbol \l},\hat{\boldsymbol \l}\}][{\boldsymbol \l},{\boldsymbol J}_1]\right)\nonumber\\
                             =&2\text{STr}\left(\{{\boldsymbol \o},[\hat{\boldsymbol \l},\{{\boldsymbol \l},\hat{\boldsymbol \l}\}]\}[{\boldsymbol \l},{\boldsymbol J}_1]\right)=0,
\end{align}
which is true due to the pure spinor condition.

For $\langle \bar b \rangle$ one needs to use the same relations from above.

\section{Conclusions and further directions}

In this paper we outlined a general method to compute the logarithmic divergences of local operators
of the pure spinor string in an $AdS_5\times S^5$ background. In the text we derived in detail the case for operators up to classical dimension two, but the method extends to any classical dimension. Although the worldsheet anomalous dimension is not related to a physical observable, as in the case of N=4 SYM, physical vertex operators should not have quantum corrections to their classical dimension. The main application of our work is to systematize the search for physical vertex operators.  We presented some consistency checks verifying that some conserved local operators are not renormalized.

The basic example is the radius operator discussed in \cite{Berkovits:2008ga}. It has ghost number $(1,1)$ and
zero classical dimension. In our notation it can be written as
\begin{align}
 V = {\rm Str}(\lambda\hat\lambda),
\end{align}
If we apply the pairing rules to compute  $\langle V\rangle$ we obtain
\begin{align}
\langle V\rangle = - {\rm I} g^{ij} {\rm Str} ( [\lambda ,T_i] [ \hat\lambda , T_j] )=0 ,
\end{align}
where in  the last  equality we replaced the structure constants and used one of the identities in the Appendix A.  This can be generalized to other massless  and massive vertex operators. We plan to return to this problem in the future.

A more interesting direction is to try to organize the dilatation operator including the higher derivative contributions. As we commented in the introduction, the difficulty here is that the pure spinor action is not an usual coset action as in \cite{Candu:2013cga,Cagnazzo:2014yha}. However, it might still be possible to obtain the complete one loop dilatation operator restricting  to some subsector of the $\mathfrak{psu}(2,2|4)$ algebra, in a way similar as it was done for super Yang-Mills dilatation operator \cite{Beisert:2003jj}.

\section*{Acknowledgements}

We would like to thank Nathan Berkovits, Osvaldo Chand\'\i a and
William Linch for useful discussion. We are especially grateful to
Luca Mazzucato for collaboration during initial stages of this work and to an anonymous Referee for several suggestions and corrections.
The work of IR is supported by CONICYT project No. 21120105 and the USM-DGIP PIIC grant. The work of BCV  is partially
supported by FONDECYT grant number 1151409.

\appendix

\section{Notation and conventions}

Here we collect the conventions and notation used in this paper. We
work with euclidean world sheet with coordinates $(z,\bar z)$.

We split the current as $J=A+K$. We define $K=J_1+J_2+J_3 \in \mathfrak{psu}(2,2|4)$ and $A=J_0$ belongs to the stability group algebra.\footnote{Although we did not use the $K$ term in the main text, it will be useful from now on to use this term in order to pack several results.} The notation that we use for the different graded generators is given by
\begin{align}
 J_0=& J_0^i T_i \quad;\quad J_1=J_1^\a T_\a \quad;\quad J_2=J_2^m T_m \quad;\quad J_3=J_3^\ha T_\ha.
\end{align}

The ghosts fields are defined as
\begin{align}
\l=&\l^A T_A\quad;\quad \o=-\o_A\eta^{A\hA}T_\hA\quad;\quad \hl=\hl^\hA T_\hA \quad;\quad \ho=\ho_\hB\eta^{B\hB}T_B.
\end{align}

The only non-zero Str of generators are
\begin{align}
 g_{ij}=&\text{STr} T_i T_j,\\
 \eta_{mn}=&\text{STr} T_m T_n,\\
 \eta_{\a\ha}=&\text{STr} T_\a T_\ha.
\end{align}

For the raising and lowering of fermionic indices in the structure constants we use
\begin{align}
 f_{m\a\b}=\eta_{\a\ha}f^\ha_{\b m} \qquad \text{and} \qquad f_{m\ha\hb}=-\eta_{\a\ha}f^\a_{\hb m},
\end{align}
and for the $f_{\a \ha i}$ the rule is the same. For the bosonic case we use the standard raising/lowering procedure.

\section{Some identities for $\mathfrak{psu}(2,2|4)$}

Let $A$, $B$ and $C$ be bosons, $X$, $Y$ and $Z$ fermions, then, the generalized Jacobi Identities are
\begin{align}
 [A,[B,C]]+[B,[C,A]]+[C,[A,B]]=&0,\\
 [A,[B,X]]+[B,[X,A]]+[X,[A,B]]=&0,\\
 \{X,[Y,A]\}+\{Y,[X,A]\}+[A,\{X,Y\}]=&0,\\
 [X,\{Y,Z\}]+[Y,\{ Z, X\}]+[Z,\{ X,Y\}]=&0.
\end{align}

In this theory the dual-coxeter number is 0, this implies
\begin{align}
 [[A,T_i],T_j]g^{ij} - \{[A,T_\a],T_\ha\}\eta^{\a\ha} +[[A,T_m],T_n]\eta^{mn} +\{[A,T_\ha],T_\a\}\eta^{\a\ha}=&0,\label{a}\\
 [[X,T_i],T_j]g^{ij} - [\{X,T_\a\},T_\ha]\eta^{\a\ha} +[[X,T_m],T_n]\eta^{mn} +[\{X,T_\ha\},T_\a] \eta^{\a\ha}=&0.\label{b}
\end{align}

The Jacobi identity yields $f_{m\a\b}f_{n\ha\hb}\eta^{mn}\eta^{\a\ha}=0$ and $f_{i\a\hb}f_{j\ha\b}g^{ij}\eta^{\a\ha}=0$. This implies that

\begin{align}
 [[J_{1,3},T_i],T_j]g^{ij}=[[J_{1,3},T_n],T_m]\eta^{mn}=
 \{[J_{1,3},T_\a],T_\ha\}\eta^{\a\ha} =\{[J_{1,3},T_\ha],T_\a\}\eta^{\a\ha}=&0,\label{c}\\
 [[\o+\l+\ho+\hl,T_i],T_j]g^{ij}=[[\o+\l+\ho+\hl,T_n],T_m]\eta^{mn}=&0,\\
 [\{\o+\l+\ho+\hl,T_\a\},T_\ha]\eta^{\a\ha}=[\{\o+\l+\ho+\hl,T_\ha\},T_\a]\eta^{\a\ha}=&0.
\end{align}

Another useful property of this theory is the pure spinor condition Eq. \ref{pureeqs}. Using it, it is easy to prove that

\begin{align}
 \left[\hl, \left[\hl, A  \right]_{\pm}\right]_{\mp}= \left[\l, \left[\l, A  \right]_{\pm}\right]_{\mp}=0.
\end{align}

\section{Complete solution of the SD equation for the $AdS_5\times S^5$ pure spinor string}

In this Appendix we apply the method explained in Section 2, and generalized in Section 3, to the $AdS_5\times S^5$ superstring. Step by step, the procedure is as follows:
\begin{enumerate}
\item Using an expansion around a classical background, $g=g_0 e^X$, we compute all the currents up to second order in $X$,
\item Expand the action (\ref{PSa}) up to second order in $X$,
\item Write down the Schwinger-Dyson equation for the model and compute the interaction matrix,
\item Compute the Green functions in powers of $\frac{1}{k}$,
\item Compute $\langle \phi^i, \phi^j\rangle$.
\end{enumerate}

The expansion of the currents was already done in (\ref{jexp}). The remaining subsections are devoted, each one, to each of the steps listed above.

We will drop the use of the boldface notation for the background fields in this section. All the quantum corrections come from either an $x$-term or a $\left(\delta\o,\delta\l,\delta\hat\o,\delta\hat\l \right)$ -term. Thus, every field in $S_{int}$, the $F$-terms, the Green's functions and in the RHS of the pairing rules should be treated as classical.

\subsection{Action}

In (\ref{exp1}) we showed the kinetic part of the expansion of (\ref{PSa}) and we promised to show the interaction part later, here we fulfil our promise. Up to second order in $X$ the interaction part is
\begingroup
\allowdisplaybreaks
\begin{align}
 S_{int}=&\frac{R^2}{2\pi}\int d^2z\left[\frac{1}{2}\bar\p x_1^\a x_1^\b J_2^m f_{m\a\b}+\frac{1}{2}x_1^\a x_1^\b J_3^{\hat \a} \bar J_3^{\hat \b}f_{i\a\hat\a}f_{j\b\hat\b}g^{ij}\right.+\frac{1}{8}\left(3x_1^\a\bar\p x_2^m-5\bar\p x_1^\a x_2^m\right)J_1^\b f_{m\a\b} \nn\\
         &+\frac{1}{8}x_1^\a x_2^m\left( -\p\bar J_1^\b f_{m\a\b}+\left[3 J_2^n \bar J_3^{\hat \a}+5\bar J_2^n J_3^{\hat \a}\right]f_{i\hat\a\a}f_{jmn}g^{ij}+3\left[J_2^n\bar J_3^{\hat \a}-\bar J_2^n J_3^{\hat\a}\right]f_{n\a\b}f_{m\hb\ha}\eta^{\b\hb}\right) \nn\\
         &-\frac{1}{4}x_1^\a x_3^{\hat\a}\left(\left[\bar J_1^\b J_3^{\hat \b}-J_1^\b \bar J_3^{\hat \b}\right]f_{m\a\b}f_{n\hat\a\hat\b}\eta^{mn}+\left[J_1^\b \bar J_3^{\hat\b} + 3\bar J_1^\b J_3^{\hat\b}\right]f_{i\hat\a\b}f_{j\a\hat\b}g^{ij}\right. \nn\\
         &\left.+J_2^m \bar J_2^n \left[f_{m\a\b}f_{n\hat\a\hb}-f_{n\a\b}f_{m\ha\hb}\right]\eta^{\b\hat\b}\right)+\frac{1}{2}\p x_3^{\hat\a}x_3^{\hat\b}\bar J_2^m f_{m\hat\a\hat\b}+\frac{1}{2}x_3^{\hat\a}x_3^{\hat\b} J_1^\a \bar J_1^\b f_{i\a\ha}f_{j\b\hat\b}g^{ij}\nn\\
         &-\frac{1}{2} x_2^m x_2^n \left(\left[J_1^\a \bar J_3^{\hat \a}-\bar J_1^\a J_3^{\hat \a}\right]f_{m\a \b}f_{n\ha \hb} \eta^{\b\hb}+J_2^p \bar J_2^q f_{ipm}f_{jqn}g^{ij}\right)+\frac{1}{8}\left(3\p x_2^m x_3^{\hat\a}\right.\nn\\
         &\left.-5x_2^m \p x_3^{\hat\a}\right)\bar J_3^{\hat \b}f_{m\hat\a\hat\b}+\frac{1}{8}x_2^m x_3^{\hat\a}\left(-\bar\p J_3^{\hat\b}f_{m\hat\a\hat\b}\right.\nn\\
         &\left.+3\left[\bar J_1^\a J_2^n-J_1^\a\bar  J_2^n \right] f_{m\a\b}f_{n\hat\a\hb}\eta^{\b\hat\b}+\left[ 3J_1^\a \bar J_2^n+5\bar J_1^\a J_2^n \right]f_{i\a\hat\a}f_{jmn}g^{ij}\right) \nn\\
         &-\d^2( N^i \hat N^j) g_{ij}-x_1^\a\left(\d N^i \bar J_3^\ha+\d\hat N^i J_3^\ha\right)f_{i\a\ha} +x_2^m\left(\d N^i \bar J_2^n+\d\hat N^i J_2^n\right)f_{imn} \nn\\
         &-x_3^\ha \left(\d N^i \bar J_1^\a+\d\hat N^i J_1^\a\right)f_{i\a\ha}-\frac{1}{2}x_1^\a x_1^\b \left(N^i \bar J_2^m + \hat N^i J_2^m\right)f_{m\a\m}f_{i\b\hm}\eta^{\m\hm} \nn\\
         &-\frac{1}{2}x_1^\a x_2^m \left( N^i \bar J_1^\b + \hat N^i J_1^\b\right)\left(f_{ipm} f_{q\a\b}\eta^{pq}+f_{i\a\hm}f_{m\b\m}\eta^{\m\hm}\right)\nn\\
         &+\frac{1}{2}\left(\p x_1^\a x_3^\ha-x_1^\a\p x_3^\ha\right) \hat N^i f_{i\a\ha}+\frac{1}{2}x_2^m \left(\bar \p x_2^n N^i+\p x_2^n \hat N^i\right)f_{imn}\nn \\
         &-\frac{1}{2}x_2^m x_3^\ha\left(N^i \bar J_3^\hb + \hat N^i J_3^\hb \right) \left(f_{ipm}f_{q\ha\hb}\eta^{pq}-f_{i\ha\m}f_{m\hb\hm}\eta^{\m\hm}\right)\nn\\
         &\left. + \frac{1}{2}x_3^\ha x_3^\hb\left( N^i \bar J_2^m + \hat N^i J_2^m\right)f_{m\ha\hm}f_{i\m\hb}\eta^{\m\hm}+\frac{1}{2}\left(\bar\p x_1^\a x_3^\ha-x_1^\a \bar\p x_3^\ha\right)N^i f_{i\a\ha}\right],
\end{align}
\endgroup
with
\begin{align}
 N^i=&-\o_A \l^B\eta^{A\hat B}f_{B\hB}^i,\\
 \hat N^i =& \ho_\hA \hl^\hB \eta^{A\hA}f_{B\hB}^i,\\
 \d N^i=&(\d\o_A \l^B+\o_A\d\l^B)\eta^{A\hat B}f_{B\hB}^i,\\
 \d\hat N^i =&( \d\ho_\hA \hl^\hB + \ho_\hA \d\hl^\hB ) \eta^{A\hA}f_{B\hB}^i,\\
 \d^2 (N^i\hat N^j)=&\d N^i \d\hat N^j-\d\o_A\d \l^B\eta^{A\hat B}f^i_{B\hB}\hat N^j+N^i\d\ho_\hA\d \hl^\hB\eta^{B\hat A}f^j_{B\hB}.
\end{align}

The lack of covariant derivatives is, as explained previously, because the pure spinor sigma model is anomaly free. This means that physical quantities only appear in gauge invariant expressions, thus the interchange $\p \leftrightarrow \P$ can be done at any moment in our computation. A more detailed explanation can be found in Subsection 3.2.

\subsection{Schwinger-Dyson equation and the Interaction Matrix}

The Schwinger-Dyson equation in momentum space for (\ref{PSa}) reads
\begingroup
\allowdisplaybreaks
\begin{align}
 G^{\a\L}=&\frac{2\pi}{R^2}\frac{\eta^{\a\L}}{|k|^2}+\frac{1}{|k|^2}(ik\bar\partial+i\bar k\partial+\Box)G^{\a\L}-\frac{\eta^{\a\Omega}}{|k|^2} F_{\Sigma\Omega} G^{\Sigma\L},\\
 G^{m\L}=&\frac{2\pi}{R^2}\frac{\eta^{m\L}}{|k|^2}+\frac{1}{|k|^2}(ik\bar\partial+i\bar k\partial+\Box)G^{\a\L}-\frac{\eta^{m\Omega}}{|k|^2} F_{\Sigma\Omega} G^{\Sigma\L},\\
 G^{\hat \a\L}=&-\frac{2\pi}{R^2}\frac{\eta^{\hat \a\L}}{|k|^2}+\frac{1}{|k|^2}(ik\bar\partial+i\bar k\partial+\Box)G^{\a\L}+\frac{\eta^{\Omega\ha}}{|k|^2} F_{\Sigma\Omega} G^{\Sigma\L},\\
 G_A^{~\L}=&\frac{2\pi}{R^2}\frac{i}{\bar k}\d_A^\L+\frac{i}{\bar k}\bar\p G_A^{~\L}-\frac{i}{\bar k}F_{\Sigma A}G^{\Sigma\L},\\
 G^{B\L}=&-\frac{2\pi}{R^2}\frac{i}{\bar k}\d_{B\L}+\frac{i}{\bar k}\bar\p G^{B\L}+\frac{i}{\bar k}F_{\Sigma }^{~B}G^{\Sigma\L},\\
 G_\hA^{~\L}=&\frac{2\pi}{R^2}\frac{i}{ k}\d_\hA^\L+\frac{i}{ k}\p G_\hA^{~\L}-\frac{i}{ k}F_{\Sigma \hA}G^{\Sigma\L},\\
 G^{\hB\L}=&-\frac{2\pi}{R^2}\frac{i}{ k}\d_{\hB\L}+\frac{i}{ k}\p G^{\hB\L}+\frac{i}{ k}F_{\Sigma }^{~\hB}G^{\Sigma\L},
\end{align}
\endgroup
where $\Lambda=\{\a,m,\hat\a,{}^A,{}_A,{}^{\hat A},{}_{\hat A}\}$.

The interaction matrix is given by

\begin{align}
 F_{\Sigma\Omega}(x,y)=&\frac{\overleftarrow{\d}}{\overleftarrow{\d}\Phi^{\Sigma}(y)}\frac{\d S_{int}}{\d \Phi^{\Omega}(x)}.
\end{align}
The directional derivative means that we compute the functional derivative of $S_{int}$ with respect to $\Phi^\Sigma$ acting from right to left.
Because we are working in momentum space is useful to write also $F$ in momentum space, for that reason the equation we work with is

\begin{align}
 F_{\L\Omega}(x,k)f(x)=&\int d^2 y \frac{\overleftarrow{\d}}{\overleftarrow{\d}\Phi^{\Sigma}(y)}\frac{\d S_{int}}{\d \Phi^{\Omega}(x)}\exp(iky)f(y).
\end{align}

Note that the $f(y)$ stands for the previous Green's function and the exponential came from the Fourier Transform. The directional derivative has the same meaning as above.

We organize the interaction matrix by the $\mathbb{Z}_4$ charge of its indices, and in the end we add the ghosts contributions.

The first we compute the $F_{\a\L}$ terms of the matrix:
\begin{align}
F_{\a\b}=&-J_2^m(i\bar k+\bar\p)f_{m\a\b} - \frac{1}{2}\bar\p J_2^m f_{m\a\b} - \frac{1}{2} J_3^{\ha}\bar J_3^{\hb} \left(f_{i\a\ha}f_{j\b\hb} - f_{i\b\ha} f_{j\a\hb}\right)g^{ij}\nonumber\\
         &+\frac{1}{2}\left(N^i \bar J_2^m + \hat N^i J_2^m\right)\left(f_{m\a\m}f_{i\b\hm}-f_{m\b\m}f_{i\a\hm}\right)\eta^{\m\hm},\\
 F_{\a m}=& J_1^\b \left(i\bar k+\bar\p\right)f_{m\a\b}+\frac{1}{8}\left(\p \bar J_1^\b+3\bar \p J_1^\b\right)f_{m\a\b} -\frac{1}{8} \left(3 J_2^n \bar J_3^{\ha} +5 \bar J_2^n J_3^{\ha}\right) f_{i\ha\a} f_{jmn}g^{ij}\\
          & -\frac{3}{8} \left(J_2^n\bar J_3^{\ha} - \bar J_2^n J_3^{\ha} \right) f_{n\a\b}f_{m\hb\ha} \eta^{\b\hb} + \frac{1}{2}\left(N^i \bar J_1^\b- \hat N^i J_1^\b\right) \left( f_{ipm} f_{q\a\b} \eta^{pq}  + f_{i\a\hm} f_{m\b\m} \eta^{\m\hm} \right) \nonumber\\
 F_{\a\ha}=& -N^i f_{i\a\ha} \left(i\bar k+\bar\p\right) - \hat N^i f_{i\a\ha} \left(i k+\p\right) + \frac{1}{4} \left(\bar J_1^\b J_3^{\hb} - J_1^\b \bar J_3^{\hb} \right) f_{m\a\b} f_{n\ha \hb} \eta^{mn} \nonumber\\
           & +\frac{1}{4}\left(J_1^\b \bar J_3^{\hat\b}+3\bar J_1^\b J_3^{\hat\b}\right)f_{i\hat\a\b}f_{j\a\hat\b}g^{ij}+\frac{1}{4}J_2^m \bar J_2^n \left(f_{m\a\b}f_{n\hat\a\hb}-f_{n\a\b}f_{m\ha\hb}\right)\eta^{\b\hat\b},\\
 F_{\a B}=&-\o_A \bar J_3^\ha A_{B~\a\ha}^A=-F_{B\a},\\
 F_\a^{~A}=&-\l^B \bar J_3^\ha A_{B~\a\ha}^A=-F^A_{~\a},\\
 F_{\a \hB}=&\ho_\hA  J_3^\ha A_{B~\a\ha}^A=-F_{\hB\a}, \\
 F_\a^{~\hA}=&\hl^\hB J_3^\ha A_{B~\a\ha}^A=-F^\hA_{~\a}.
\end{align}

The terms of the $F_{m\L}$ kind are
\begingroup
\allowdisplaybreaks
\begin{align}
 F_{m\a}=& J_1^\b \left(i\bar k+\bar\p\right) f_{m\a\b}  + \frac{1}{2} \left( N^i \bar J_1^\b + \hat N^i J_1^\b \right) \left(f_{ipm}f_{q\a\b} \eta^{pq} + f_{i\a\hm}f_{m\b\m} \eta^{\m\hm}\right)\nonumber\\
         & +\frac{1}{8} \left( 3 J_2^n \bar J_3^{\ha} + 5\bar J_2^n J_3^{\ha} \right)f_{i\ha\a} f_{jmn} g^{ij} + \frac{3}{8}\left(J_2^n\bar J_3^{\hat \a}-\bar J_2^n J_3^{\hat\a}\right) f_{n\a\b}f_{m\hb\ha}\eta^{\b\hb}\nonumber\\& + \frac{1}{8}\left(5\bar \p J_1^\b - \p\bar J_1^\b\right) f_{m\a\b},\\
 F_{mn}=& N^if_{imn}\left(i\bar k+\bar\p\right)+\hat N^if_{imn}\left(ik+\p\right)\nonumber \\
        &-\frac{1}{2} \left(J_1^\a \bar J_3^{\hat \a}-\bar J_1^\a J_3^{\hat \a}\right)(f_{m\a \b}f_{n\ha \hb}+f_{n\a \b}f_{m\ha \hb}) \eta^{\b\hb}-\frac{1}{2}J_2^p \bar J_2^q (f_{ipm}f_{jqn}+f_{ipn}f_{jqm})g^{ij},\\
 F_{m\ha}=&\bar J_3^\hb f_{m\ha\hb}\left(ik+\p\right) + \frac{1}{8}\left(5\p\bar J_3^\hb-\bar \p J_3^\hb\right)f_{m\ha\hb}+\frac{3}{8}\left( \bar J_1^\a J_2^n-J_1^\a\bar  J_2^n \right) f_{m\a\b}f_{n\hat\a\hb} \eta^{\b\hb}\\
          &+\frac{1}{8}\left( 3J_1^\a \bar J_2^n+5\bar J_1^\a J_2^n \right)f_{i\a\hat\a}f_{jmn}g^{ij} + \frac{1}{2}\left(N^i \bar J_3^\hb + \hat N^i J_3^\hb\right)\left(f_{ipm}f_{q\ha\hb}\eta^{pq} -f_{i\ha\m}f_{m\hb\hm} \eta^{\m\hm}\right),\nonumber\\
 F_{mB}=&-\o_A \bar J_2^n A_{B~mn}^A=F_{Bm},\\
 F_{m}^{~A}=&-\l^B \bar J_2^n A_{B~mn}^A=F_{~m}^B,\\
 F_{m\hB}=&\ho_\hA  J_2^n A_{B~mn}^A=F_{\hB m},\\
 F_{m}^{~\hA}=&\l^\hB J_2^n A_{B~mn}^A=F_{~m}^\hA.
\end{align}
\endgroup

The last contribution from the non-ghost terms is given by the $F_{\ha \L}$ elements:
\begin{align}
 F_{\ha\a}=& -N^i f_{i\a\ha}\left(i\bar k+\bar\p\right)-\hat N^i f_{i\a\ha}\left(i k+\p\right) -\frac{1}{4}\left( J_1^\b J_3^{\hat \b}-J_1^\b \bar J_3^{\hat \b}\right)f_{m\a\b}f_{n\hat\a\hat\b}\eta^{mn}\nonumber \\&-\frac{1}{4}\left(J_1^\b \bar J_3^{\hat\b}+3\bar J_1^\b J_3^{\hat\b}\right)f_{i\hat\a\b}f_{j\a\hat\b}g^{ij}-\frac{1}{4}J_2^m \bar J_2^n \left(f_{m\a\b}f_{n\hat\a\hb}-f_{n\a\b}f_{m\ha\hb}\right)\eta^{\b\hat\b},\\
 F_{\ha m}=&\bar J_3^\hb f_{m\ha\hb}\left(ik+\p\right) + \frac{1}{8} \left(3\p\bar J_3^\hb+\bar\p J_3^\hb\right)f_{m\ha\hb}-\frac{3}{8}\left(\bar J_1^\a J_2^n-J_1^\a\bar  J_2^n \right)f_{m\a\b}f_{n\hat\a\hb}\eta^{\b\hb}\\
          &-\frac{1}{8}\left( 3J_1^\a \bar J_2^n+5\bar J_1^\a J_2^n \right)f_{i\a\hat\a}f_{jmn}g^{ij} - \frac{1}{2}\left(N^i \bar J_3^\hb + \hat N^i J_3^\hb\right)\left(f_{ipm}f_{q\ha\hb}\eta^{pq} -f_{i\ha\m}f_{m\hb\hm}\eta^{\m\hm}\right),\nonumber\\
 F_{\ha\hb}=&-\bar J_2^m \left(ik+\p\right)f_{m\ha\hb}-\frac{1}{2}\p\bar J_2^m f_{m\ha\hb}-\frac{1}{2} J_1^\a \bar J_1^\b\left(   f_{i\a\ha}f_{j\b\hat\b}-f_{i\b\ha}f_{j\a\hb}\right)g^{ij}\nonumber\\
           & - \frac{1}{2}\left( N^i \bar J_2^m + \hat N^i J_2^m\right)\left( f_{m\ha\hm}f_{i\m\hb}-f_{m\hb\hm}f_{i\m\ha} \right) \eta^{\m\hm},\\
 F_{\ha B}=&-\o_A \bar J_1^\a A_{B~\a\ha}^A=-F_{B\ha},\\
 F_{\ha }^{~A}=&-\l^B \bar J_1^\a A_{B~\a\ha}^A=-F_{~\ha}^A,\\
 F_{\ha \hB}=&\ho_\hA  J_1^\a A_{B~\a\ha}^A=-F_{\hB\ha},\\
 F_{\ha B}=&\l^\hA J_1^\a A_{B~\a\ha}^A=-F_{~\ha}^\hB.
\end{align}

Finally we compute the pure ghost terms, and we save some trees by not adding the symmetric terms already listed:
\begingroup
\allowdisplaybreaks
\begin{align}
 F_{B}^{~A}=&\hat N^B_A=F_{~B}^{A},\\
 F_{B}^{~\hA}=&\o_A \hl^\hB A_{B\hB}^{A\hA}=F_{~B}^{\hA},\\
 F_{B\hB}=&\o_A\ho_\hA A_{B\hB}^{A\hA}=F_{B\hA},\\
 F_{~\hB}^{A}=&\l^B \ho_\hA A_{B\hB}^{A\hA}=F_{\hB}^{~A},\\
 F^{A\hA}=&\l^B\hl^\hB A_{B\hB}^{A\hA}=F^{\hA A},\\
 F_\hB^{~\hA}=&N_\hB^\hA=F^\hA_{~\hB},
\end{align}
\endgroup
where we have defined
\begin{align}
 A_{B\hB}^{A\hA}=&\eta^{A\hC}\eta^{C\hA}f^i_{B\hC}f^j_{\hB C}g_{ij},\\
 \hat N^B_A=&\ho_\hA \hl^\hB A_{B\hB}^{A\hA},\\
 N^\hB_\hA=&\o_A \l^B A_{B\hB}^{A\hA}.
\end{align}

\subsection{Green functions}

With all the previous results, we begin the computation of the Green's Functions as a power series in $1/k$. We follow the prescription given in (\ref{green}). The Green functions are presented order by order, which makes the reading easier.

The only contributions of order $1/k$ come from the ghosts propagators
\begin{align}
 G_{1A}^{~~B}=&\frac{2\pi}{R^2}\frac{i}{\bar k}\d^B_A=-G_{1~A}^B, \\
 G_{1\hA}^{~~\hB}=&\frac{2\pi}{R^2}\frac{i}{k}\d^\hB_\hA=-G_{1~\hA}^\hB.
\end{align}

For the $1/k^2$ terms, we have a contribution from the non-ghosts propagators
\begin{align}
 G_{2}^{\a\ha}=&\frac{2\pi}{R^2}\frac{1}{|k|^2}\eta^{\a\ha}=-G_2^{\ha\a}, \\
 G_2^{mn}=&\frac{2\pi}{R^2}\frac{1}{|k|^2}\eta^{mn},
\end{align}
and another from the ghost interactions
\begingroup
\allowdisplaybreaks
\begin{align}
 G_{2A}^{~~B}=&-\frac{i}{\bar k}\left(F_{A}^{~C}G_{1C}^{~~B}\right)
             =\frac{2\pi}{R^2}\frac{1}{\bar k^2}\hat N_A^B=G_{2~A}^B,\\
 G_{2A\hA}=&-\frac{i}{\bar k}\left(F_{A\hat C}G_{1~\hA}^{\hat C}\right)
          =-\frac{2\pi}{R^2}\frac{1}{|k|^2}\o_B\ho_\hB A_{A\hA}^{B\hB}=G_{2\hA A},\\
 G_{2A}^{~~\hB}=&-\frac{i}{\bar k}\left(F_{A}^{~\hat C}G_{1\hat C}^{~~\hB}\right)
               =\frac{2\pi}{R^2}\frac{1}{|k|^2}\o_B\hl^\hA A_{A\hA}^{B\hB}=G_{2~ A}^\hB,\\
 G_{2~\hA}^B=&\frac{i}{\bar k}\left(F^B_{~\hat C}G_{1~~\hA}^{\hat C}\right)
            =\frac{2\pi}{R^2}\frac{1}{|k|^2}\l^A\ho_\hB A_{A\hA}^{B\hB}=G_{2\hA}^B,\\
 G_{2}^{B\hB}=&\frac{i}{\bar k}\left(F^{B\hat C}G_{1\hat C}^{~~\hat B}\right)
             =-\frac{2\pi}{R^2}\frac{1}{|k|^2}\l^A\hl^\hA A_{A\hA}^{B\hB}=G_2^{\hB B},\\
 G_{2\hA}^{~~\hB}=&-\frac{i}{k}\left(F_{\hA}^{~\hat C}G_{1\hat C}^{~~\hB}\right)
                 =\frac{2\pi}{R^2}\frac{1}{k^2}N^\hB_\hA=G_{2~\hA}^\hB.
\end{align}
\endgroup

At order $1/k^3$ we have interaction between the non-ghost fields. We organize these terms in the same order as in the previous section, when $G_{\L\Omega}=c G_{\Omega\L}$, with $c=\pm 1$ we only list the first term.

Using the given prescription, we find that the $G_{3}^{\a\L}$ terms are
\begin{align}
 G_3^{\a\b}=&-\frac{\eta^{\a\ha}}{|k|^2}\left(F_{\hb\ha}G_2^{\hb\b}\right)
           =-\frac{2\pi}{R^2}\frac{i}{|k|^2}\frac{\bar J_2^m}{\bar k}f_{m\ha\hb}\eta^{\a\ha}\eta^{\b\hb},\\
 G_3^{\a m}=&-\frac{\eta^{\a\ha}}{|k|^2}\left(F_{n\ha}G_2^{nm}\right)
           =-\frac{2\pi}{R^2}\frac{i}{|k|^2}\frac{\bar J_3^\hb}{\bar k}f_{n\ha\hb}\eta^{\a\ha}\eta^{mn}=-G_3^{m\a},\\
 G_3^{\a\ha}=&-\frac{\eta^{\a\hb}}{|k|^2}\left(F_{\b\hb}G_2^{\b\ha}\right)
           =\frac{2\pi}{R^2}\frac{i}{|k|^2}\left(\frac{N^i}{k}+\frac{\hat N^i}{\bar k}\right)f_{i\b\hb}\eta^{\a\hb}\eta^{\b\ha}=G_3^{\ha\a},\\
 G_{3~A}^{\a}=&-\frac{\eta^{\a\ha}}{|k|^2}\left(F_{B\ha}G_{1~A}^B\right)
             =\frac{2\pi}{R^2}\frac{i}{|k|^2}\frac{\bar J_1^\b}{\bar k}\o_B A_{A~\b\ha}^{~B}\eta^{\a\ha}=-G_{3A}^{~~\a},\\
 G_3^{\a B}=&-\frac{\eta^{\a\ha}}{|k|^2}\left(F_{~\ha}^A G_{1A}^{~~B}\right)
           =-\frac{2\pi}{R^2}\frac{i}{|k|^2}\frac{\bar J_1^\b}{\bar k}\l^A A_{A~\b\ha}^{~B}\eta^{\a\ha}=-G_{3}^{B\a},\\
 G_{3~\hA}^{\a}=&-\frac{\eta^{\a\ha}}{|k|^2}\left(F_{\hB\ha}G_{1~\hA}^\hB\right)
             =-\frac{2\pi}{R^2}\frac{i}{|k|^2}\frac{J_1^\b}{k}\ho_\hB A_{\hA~\b\ha}^{~\hB}\eta^{\a\ha}=-G_{3\hA}^{~~\a},\\
 G_3^{\a \hB}=&-\frac{\eta^{\a\ha}}{|k|^2}\left(F_{~\ha}^\hA G_{1\hA}^{~~\hB}\right)
           =\frac{2\pi}{R^2}\frac{i}{|k|^2}\frac{J_1^\b}{k}\hl^\hA A_{\hA~\b\ha}^{~\hB}\eta^{\a\ha}=-G_{3}^{\hB\a}.
\end{align}

For the $G_{3}^{m\L}$ terms we find
\begin{align}
 G_3^{mn}=&-\frac{\eta^{mp}}{|k|^2}\left(F_{q p} G_2^{qn}\right)
         =-\frac{2\pi}{R^2}\frac{i}{|k|^2}\left(\frac{N^i}{k}+\frac{\hat N^i}{\bar k}\right)f_{ipq}\eta^{mp}\eta^{nq},\\
 G_3^{m\ha}=&-\frac{\eta^{mn}}{|k|^2}\left(F_{\a n} G_2^{\a\ha}\right)
           =-\frac{2\pi}{R^2}\frac{i}{|k|^2}\frac{J_1^\b}{k}f_{n\a\b}\eta^{\a\ha}\eta^{mn}=-G_3^{\ha m}\\
 G_{3~A}^{m}=&-\frac{\eta^{mn}}{|k|^2}\left(F_{Bn}G_{1~A}^B\right)
            =-\frac{2\pi}{R^2}\frac{i}{|k|^2}\frac{\bar J_2^p}{\bar k}\o_B A_{A~np}^{~B}\eta^{mn}=-G_{3A}^{~~m},\\
 G_3^{m B}=&-\frac{\eta^{mn}}{|k|^2}\left(F_{~\ha}^A G_{1A}^{~~B}\right)
          =\frac{2\pi}{R^2}\frac{i}{|k|^2}\frac{\bar J_2^p}{\bar k}\l^A A_{A~np}^{~B}\eta^{mn}=G_{3}^{Bm},\\
 G_{3~\hA}^{m}=&-\frac{\eta^{mn}}{|k|^2}\left(F_{\hB\ha}G_{1~\hA}^\hB\right)
              =\frac{2\pi}{R^2}\frac{i}{|k|^2}\frac{J_2^p}{k}\ho_\hB A_{\hA~np}^{~\hB}\eta^{mn}=-G_{3\hA}^{~~m},\\
 G_3^{m \hB}=&-\frac{\eta^{mn}}{|k|^2}\left(F_{~\ha}^\hA G_{1\hA}^{~~\hB}\right)
            =-\frac{2\pi}{R^2}\frac{i}{|k|^2}\frac{J_2^p}{k}\hl^\hA A_{\hA~np}^{~\hB}\eta^{mn}=-G_{3}^{\hB m}.
\end{align}

The $G_{3}^{\ha\L}$ terms computed are
\begingroup
\allowdisplaybreaks
\begin{align}
 G_3^{\ha\hb}=&-\frac{2\pi}{R^2}\frac{i}{|k|^2}\frac{J^m_2}{k}f_{m\a\b}\eta^{\a\ha}\eta^{\b\hb}\\
 G_{3~A}^{\ha}=&\frac{\eta^{\a\ha}}{|k|^2}\left(F_{B\ha}G_{1~A}^B\right)
              =-\frac{2\pi}{R^2}\frac{i}{|k|^2}\frac{\bar J_3^\hb}{\bar k}\o_B A_{A~\hb\a}^{~B}\eta^{\a\ha}=-G_{3A}^{~~\ha},\\
 G_3^{\ha B}=&\frac{\eta^{\a\ha}}{|k|^2}\left(F_{~\ha}^A G_{1A}^{~~B}\right)
            =\frac{2\pi}{R^2}\frac{i}{|k|^2}\frac{\bar J_3^\hb}{\bar k}\l^A A_{A~\hb\a}^{~B}\eta^{\a\ha}=-G_{3}^{B\ha},\\
 G_{3~\hA}^{\ha}=&\frac{\eta^{\a\ha}}{|k|^2}\left(F_{\hB\ha}G_{1~\hA}^\hB\right)
                =\frac{2\pi}{R^2}\frac{i}{|k|^2}\frac{J_3^\hb}{k}\ho_\hB A_{\hA~\hb\a}^{~\hB}\eta^{\a\ha}=-G_{3\hA}^{~~\ha},\\
 G_3^{\ha \hB}=&\frac{\eta^{\a\ha}}{|k|^2}\left(F_{~\ha}^\hA G_{1\hA}^{~~\hB}\right)
              =-\frac{2\pi}{R^2}\frac{i}{|k|^2}\frac{J_3^\hb}{k}\hl^\hA A_{\hA~\hb\a}^{~\hB}\eta^{\a\ha}=-G_{3}^{\hB\ha},\\
\end{align}
\endgroup

The $G_3$ with only ghost indices are
\begin{align}
 G_{3AC}=&-\frac{2\pi}{R^2}\frac{i}{|k|^2}\frac{1}{\bar k}\o_B\o_D\hl^\hA \ho_\hB\left[A_{A\hA}^{B\hat C}A_{C\hat C}^{D\hB}-A_{A\hat C}^{B\hB}A_{C\hA}^{D\hat C}\right],\\
 {G_{3A}}^B=&\frac{2\pi}{R^2}\frac{i}{\bar k^3}\left(\d_A^D\bar \p - \hat N^D_A\right) \hat N_D^B +\frac{2\pi}{R^2}\frac{i}{|k|^2}\frac{1}{\bar k}\o_D\l^C \hl^\hA \ho_\hB \left[A_{A\hat C}^{D\hat B}A_{C\hA}^{B\hat C}-A_{A\hA}^{D\hat C}A_{C\hat C}^{B\hB}\right],\\
 G_{3A\hA}=&-\frac{2\pi}{R^2}\frac{i}{|k|^2}\frac{1}{\bar k}\left(\d_A^D\bar \p -\hat N^D_A\right)\o_B\ho_\hB A_{D\hA}^{B\hB}-\frac{2\pi}{R^2}\frac{i}{|k|^2}\frac{1}{k} \o_B\ho_\hB N^{\hat D}_{\hA} A_{A\hat D}^{B\hB},\\
 G_{3A}^{~~\hB}=&\frac{2\pi}{R^2}\frac{i}{|k|^2}\frac{1}{\bar k}\left(\d_A^D\bar \p - \hat N^D_A\right)\o_B\hl^\hA A_{D\hA}^{B\hB}-\frac{2\pi}{R^2}\frac{i}{|k|^2}\frac{1}{k}\o_B\hl^\hA N^{\hB}_{\hat D} A_{A\hA}^{B\hat D},\\
 {G_3^B}_A=&\frac{2\pi}{R^2}\frac{i}{\bar k^3}\left(\d_D^B\bar \p +\hat N^B_D \right)\hat N_A^D + \frac{2\pi}{R^2}\frac{i}{|k|^2}\frac{1}{\bar k}\o_D\l^C \hl^\hA \ho_\hB \left[A_{A\hA}^{D\hC}A_{C\hC}^{B\hB} - A_{A\hC}^{D\hB}A_{C\hA}^{B\hC}\right], \\
 G_3^{BD}=&\frac{2\pi}{R^2}\frac{i}{|k|^2}\frac{1}{\bar k}\l^A \l^C \hl^\hA \ho_\hB \left[A_{A\hA}^{B\hC}A_{C\hC}^{D\hB} - A_{A\hC}^{B\hB}A_{C\hA}^{D\hC}\right],\\
 G_{3~\hA}^B=&\frac{2\pi}{R^2}\frac{i}{\bar k}\frac{1}{|k|^2}\left(\d_D^B\bar \p +\hat N^B_D \right)\l^A\ho_\hB A_{A\hA}^{B\hB}+\frac{2\pi}{R^2}\frac{i}{|k|^2}\frac{1}{k}\l^A \ho_\hB N^{\hat D}_\hA A_{A\hat D}^{B\hB},\\
 G_3^{B\hB}=&-\frac{2\pi}{R^2}\frac{i}{\bar k}\frac{1}{|k|^2}\left(\d_D^B\bar \p +\hat N^B_D \right)\l^A\hl^\hA A_{A\hA}^{B\hB}+\frac{2\pi}{R^2}\frac{i}{|k|^2}\frac{1}{k}\l^A \hl^\hA N^{\hB}_\hD A_{A\hA}^{B\hD},\\
 G_{3\hA A}=&\frac{2\pi}{R^2}\frac{i}{|k|^2}\frac{1}{k}\left(-\d^\hD_\hA\p +N^\hD_\hA\right)\o_B \ho_\hB A_{A\hA}^{B\hB} -\frac{2\pi}{R^2}\frac{i}{|k|^2}\frac{1}{\bar k}\o_B\ho_\hB \hat N^D_A A_{D\hA}^{B\hB},\\
 {G_{3\hA}}^B=&-\frac{2\pi}{R^2}\frac{i}{|k|^2}\frac{1}{k}\left(-\d^\hD_\hA\p +N^\hD_\hA\right)\l^A \ho_\hB A_{A\hA}^{B\hB} -\frac{2\pi}{R^2}\frac{i}{|k|^2}\frac{1}{\bar k}\l^A\ho_\hB \hat N^B_C A_{A\hA}^{C\hB},\\
 G_{3\hA\hat C}=&\frac{2\pi}{R^2}\frac{i}{|k|^2}\frac{1}{k}\ho_{\hat B}\ho_{\hat D}\l^A\o_B\left[A_{A\hA}^{C\hB}A_{C\hC}^{B\hD}-A_{C\hA}^{B\hB}A_{A\hC}^{C\hD}\right],\\
 {G_{3\hA}}^\hB=&-\frac{2\pi}{R^2}\frac{i}{k^3}\left(-\d^\hD_\hA\p +N^\hD_\hA\right)N_\hD^\hB + \frac{2\pi}{R^2}\frac{i}{|k|^2}\frac{1}{k}\ho_{\hat D}\hl^{\hat C} \l^A\o_B\left[A_{C\hA}^{B\hD}A_{A\hC}^{C\hB}-A_{A\hA}^{C\hD}A_{C\hC}^{B\hB}\right],
\end{align}

\begin{align}
 {G_3^\hB}_A=&\frac{2\pi}{R^2}\frac{i}{|k|^2}\frac{1}{k}\left(\d^\hB_\hD\p + N^\hB_\hD\right)\o_B \hl^\hA A_{A\hA}^{B\hD} + \frac{2\pi}{R^2}\frac{i}{|k|^2}\frac{1}{\bar k}\o_B\hl^\hA \hat N^C_A A_{C\hA}^{B\hB},\\
 G_3^{\hB B}=&-\frac{2\pi}{R^2}\frac{i}{|k|^2}\frac{1}{k}\left(\d^\hB_\hD\p + N^\hB_\hD\right)\l^A \hl^\hA A_{A\hA}^{B\hD} + \frac{2\pi}{R^2}\frac{i}{|k|^2}\frac{1}{\bar k}\l^A \hl^\hA \hat N_A^B A_{A\hA}^{C\hB},\\
 {G_3^\hB}_\hA=&\frac{2\pi}{R^2}\frac{i}{k^3}\left(\d^\hB_\hD\p + N^\hB_\hD\right)N^\hD_\hA - \frac{2\pi}{R^2}\frac{i}{|k|^2}\frac{1}{k}\ho_{\hat D}\hl^{\hat C} \l^A\o_B\left[A_{A\hC}^{C\hB}A_{C\hA}^{B\hD}-A_{C\hC}^{B\hB}A_{A\hA}^{C\hB}\right],\\
 G_3^{\hat B\hat D}=&\frac{2\pi}{R^2}\frac{i}{|k|^2}\frac{1}{k}\hl^\hA \hl^\hC \l^A\o_B\left[A_{A\hA}^{C\hB}A_{C\hC}^{B\hD}-A_{C\hA}^{B\hB}A_{A\hC}^{C\hD}\right].
\end{align}

The $1/k^4$ terms are needed when we compute terms with two derivatives. Since we are not computing anything with two derivatives and at least one ghost field, we don't list those Green's functions. The $G_4^{\a\L}$ terms are:
\begin{align}
 G_4^{\a\b}=&\frac{2\pi}{R^2}\frac{1}{|k|^2\bar k^2}\left(\bar\p \bar J_2^m f_{m\ha\hb}+\bar J_2^m \hat N^i\left[ f_{i\m\ha}f_{m\hm\hb}-f_{i\m\hb}f_{m\hm\ha}\right]\eta^{\m\hm}+\bar J_3^\hm\bar J_3^\hn f_{m\ha\hm}f_{n\hb\hn}\eta^{mn}\right)\eta^{\a\ha}\eta^{\b\hb}\nonumber\\
            &+\frac{2\pi}{R^2}\frac{1}{|k|^4}\left(\frac{1}{2}\p \bar J_2^m f_{m\ha\hb}+\frac{1}{2}J_1^\m\bar J_1^\n g^{ij}\left(f_{i\m\ha}f_{j\n\hb}-f_{i\n\ha}f_{j\m\hb}\right)\right.\nonumber\\
            &\left. + \frac{1}{2}\left[-\bar J_2^m N^i+J_2^m\hat N^i\right]\left(f_{m\ha\hm}f_{i\m\hb}-f_{m\hb\hm}f_{i\m\ha}\right)\eta^{\m\hm}\right)\eta^{\a\ha}\eta^{\b\hb},\\
 G_4^{\a m}=&\frac{2\pi}{R^2}\frac{1}{|k|^2\bar k^2}\left[\bar \p\bar J_3^\hb f_{n\ha\hb}+\bar J_3^\hb \hat  N^i\left(f_{p\ha\hb}f_{inq}\eta^{pq}+f_{n\hm\hb}f_{i\m\ha}\eta^{\m\hm}\right)\right]\eta^{mn}\eta^{\a\ha}\nonumber\\
            &+\frac{2\pi}{R^2}\frac{1}{|k|^4}\left[\frac{1}{8}\left(3 \p\bar J_3^\hb+\bar \p J_3^\hb\right) f_{n\ha\hb}-\frac{1}{2}(3N^i \bar J_3^\hb+\hat N^i J_3^\hb)\left(f_{ipn}f_{q\ha\hb}\eta^{pq}-f_{i\ha\m}f_{n\hb\hm}\eta^{\m\hm}\right)\right.\nonumber\\
            &\left.-\frac{1}{8}\left[5\bar J_1^\b J_2^p+3 J_1^\b \bar J_2^p \right]f_{i\b\ha}f_{jnp}g^{ij}-\frac{1}{8}\left[3\bar J_1^\b J_2^p+5J_1^\b\bar J_2^p\right]f_{n\b\m}f_{p\ha\hm}\eta^{\m\hm}\right]\eta^{\a\ha}\eta^{mn},\\
 G_4^{\a\ha}=&\frac{2\pi}{R^2}\frac{1}{|k|^2 k^2}\left[-\p N^i f_{i\b\hb}-N^i N^j f_{i\m\hb}f_{j\b\hm}\eta^{\m\hm}\right] \eta^{\a\hb}\eta^{\b\ha} \nonumber\\
             &+ \frac{2\pi}{R^2}\frac{1}{|k|^2 \bar k^2}\left[-\bar \p \hat N^i f_{i\b\hb}-\hat N^i\hat  N^j f_{i\m\hb}f_{j\b\hm}\eta^{\m\hm}\right]\eta^{\a\hb}\eta^{\b\ha}\nonumber\\
             &+\frac{2\pi}{R^2}\frac{1}{|k|^4}\left[-\left(N^i\hat N^j+N^j\hat N^i\right)f_{i\m\hb}f_{j\hm\b}\eta^{\m\hm}+\frac{1}{4}J_2^m \bar J_2^n\left(3f_{m\m\b}f_{n\hb\hm}+f_{n\m\b}f_{m\hb\hm}\right)\eta^{\m\hm}  \right.\nonumber\\
             &\left.+\frac{1}{4}J_1^\m \bar J_3^\hm\left(5f_{m\m\b}f_{n\hm\hb}\eta^{mn}-f_{i\m\hb} f_{j\hm\b} g^{ij} \right)-\frac{1}{4}\bar J_1^\m J_3^\hm\left( f_{m\m\b}f_{n\hm\hb}\eta^{mn} +3f_{i\m\hb} f_{j\hm\b} g^{ij} \right) \right] \eta^{\a\hb} \eta^{\b\ha}.
\end{align}

The $G_4^{m\L}$ Green's functions are
\begingroup
\allowdisplaybreaks
\begin{align}
 G_4^{mn}=&\frac{2\pi}{R^2}\frac{1}{|k|^2 \bar k^2}\left[\bar \p\hat N^i f_{ipq}- \hat N^i\hat N^j f_{irp}f_{jsq}\eta^{rs}\right]\eta^{nq}\eta^{mp}\nonumber\\
          & + \frac{2\pi}{R^2}\frac{1}{|k|^2 k^2}\left[ \p N^i f_{ipq} - N^i N^j f_{irp}f_{jsq}\eta^{rs}\right]\eta^{nq}\eta^{mp}\nonumber\\
          &+\frac{2\pi}{R^2}\frac{1}{|k|^4}\eta^{mp}\eta^{nq}\left[ - \left(N^i \hat N^j+N^j\hat N^i\right)f_{irp}f_{jsq}\eta^{rs}+\frac{1}{2}J_2^r \bar J_2^s \left(f_{irp}f_{jsq}+f_{irq}f_{jsp}\right)g^{ij}\right.\nonumber\\
          &\left.-\frac{1}{2}\left(J_1^\a \bar J_3^\ha +\bar J_1^\a J_3^\ha \right)\left( f_{q\a\b}f_{p\ha\hb}+f_{p\a\b}f_{q\ha\hb} \right)\eta^{\b\hb}\right],\\
 G_4^{\ha m}=&\frac{2\pi}{R^2}\frac{1}{|k|^2 k^2}\left[-\p J_1^\b f_{n\a\b} + J_1^\b N^i \left( f_{ipn}f_{q\a\b}\eta^{pq} + f_{n\m\b} f_{i\hm\a} \eta^{\m\hm} \right)\right]\eta^{mn}\eta^{\a\ha}\nonumber\\
            &+\frac{2\pi}{R^2}\frac{1}{|k|^4}\left[-\frac{1}{8}\left(3\bar \p J_1^\b+\p \bar J_1^\b\right) f_{n\a\b}+\frac{1}{2}\left( N^i \bar J_1^\b + 3\hat N^i J_1^\b\right) \left(f_{ipn}f_{q\a\b}\eta^{pq}+f_{i\a\hm}f_{n\b\m}\eta^{\m\hm}\right)\right.\nonumber\\
             &\left.+\frac{1}{8}\left(3 J_2^p \bar J_3^\hb+5 \bar J_2^p J_3^\hb\right)f_{i\a\hb}f_{jnp}g^{ij}-\frac{1}{8}\left(5 J_2^p \bar J_3^\hb + 3 \bar J_2^p J_3^\hb\right)f_{p\a\m}f_{n\hb\hm}\eta^{\m\hm}\right]\eta^{\a\ha}\eta^{nm}.
\end{align}
\endgroup

Finally, we list the $G_4^{\ha\hb}$ term
\begin{align}
\nonumber G_4^{\ha\hb}=&\frac{2\pi}{R^2}\frac{1}{|k|^2 k^2}\left[\p J_2^m f_{m\a\b} + J_2^m N^i\left(f_{m\a\m}f_{i\b\hm}-f_{m\b\m}f_{i\a\hm}\right)\eta^{\m\hm}+ J_1^\m J_1^\n f_{m\a\m}f_{n\b\n}\eta^{mn}\right]\eta^{\a\ha}\eta^{\b\hb}\\
\nonumber              &+\frac{2\pi}{R^2}\frac{1}{|k|^4}\left[\frac{1}{2}\bar \p J_2^m f_{m\a\b}+\frac{1}{2}J_3^\hm \bar J_3^\hn\left(f_{i\hm\a}f_{j\hn\b}-f_{i\hm\b}f_{j\hn\a}g^{ij}\right)\right.\\
              &\left.+ \frac{1}{2}\left(N^i\bar J_2^m - \hat N^i J_2^m\right)\left(f_{m\b\m}f_{i\a\hm}-f_{m\a\m}f_{i\b\hm}\right)\eta^{\m\hm}\right]\eta^{\a\ha}\eta^{\b\hb}.
\end{align}

The reason we don't compute terms such as $G_4^{\ha m}$ is that we can deduce their contribution from the relation $\langle \p X \p X \rangle=\p \langle X \p X \rangle - \langle X\p\p X \rangle$, as explained in section 2.

\subsection{Pairing rules}
We split the current in its gauge part $J_0$ and the vielbein $K$:
\begin{align}
 J=& J_0+K,\\
 K=&J_1+J_2+J_3.
\end{align}

We also join the quantum fluctuations into a single term
\begin{align}
 X=x_1+x_2+x_3.
\end{align}

The following is the list of all divergent parts up to two derivatives. The order of the results is: first terms with no derivatives, then the currents, then one $X$ with one current, and finally two currents. Finally, we list the pairing rules involving ghost fields. The definition of ${\rm I}$ in this appendix is ${\rm I}=-{1/(2R^2 \e)} $.

The non-vanishing terms with no derivatives are the ones given by the first term in the Schwinger-Dyson equation:
\begin{align}
 \langle x_1,x_3\rangle = -T_\a T_\ha \eta^{\a\ha} \qquad \text{and} \qquad \langle x_2,x_2\rangle=T_mT_n \eta^{mn}. \label{xx}
\end{align}

Now we show the divergent part of the currents:
\begin{align}
 \langle K \rangle=&\langle \bar K \rangle=\langle N \rangle=\langle \hat N \rangle=0,\label{j}\\
 \langle J_0 \rangle=&-\frac{{\rm I}}{2}\left(\{[N,T_\ha],T_\a \}\eta^{\a\ha}-\{[N,T_\a],T_\ha\}\eta^{\a\ha}+[[N,T_m],T_n]\eta^{mn}\right),\label{<j0>}\\
 \langle \bar J_0 \rangle=&-\frac{\rm I}{2}\left(\{[\hat N,T_\ha],T_\a \}\eta^{\a\ha}-\{[\hat N,T_\a],T_\ha\}\eta^{\a\ha}+[[\hat N,T_m],T_n]\eta^{mn}\right)\label{<j0b>}.
\end{align}

For one $X$ with one current, we find that the simplest current is $J_0$
\begin{align}
 \langle X,J_0 \rangle=&-{\rm I}[K,T_j]T_k g^{jk},\\
 \langle X, \bar J_0 \rangle=&-{\rm I}[\bar K,T_j]T_k g^{jk},
\end{align}
for the other currents we find
\begin{align}
 \langle x_1,J_1 \rangle=&-{\rm I}[J_2,T_\ha]T_\a\eta^{\a\ha}, &&  \langle x_2,J_1 \rangle={\rm I}[J_3,T_\ha]T_\a\eta^{\a\ha}, &  \langle x_3,J_1 \rangle=&{\rm I}[N,T_\ha]T_\a\eta^{\a\ha}, \label{xj1}\\
 \langle x_1,\bar J_1 \rangle=&0, &&  \langle x_2,\bar J_1 \rangle=0,  &  \langle x_3,\bar J_1 \rangle=&{\rm I}[\hat N,T_\ha]T_\a\eta^{\a\ha},\label{xj1b}\\
 \langle x_1,J_2 \rangle=&-{\rm I}[J_3,T_m]T_n\eta^{mn}, &&  \langle x_2,J_2 \rangle={\rm I}[N,T_m]T_n\eta^{mn}, &  \langle x_3,J_2 \rangle=&0,\label{xj2}\\
 \langle x_1,\bar J_2 \rangle=&0, && \langle x_2,\bar J_2 \rangle={\rm I}[\hat N,T_m]T_n\eta^{mn}, &  \langle x_3,\bar J_2 \rangle=&{\rm I}[\bar J_1,T_m]T_n\eta^{mn},\label{xj2b}\\
 \langle x_1,J_3 \rangle=&-{\rm I}[N,T_\a]T_\ha\eta^{\a\ha}, &&  \langle x_2,J_3 \rangle=0, &  \langle x_3,J_3 \rangle=&0,\label{xj3}\\
 \langle x_1,\bar J_3 \rangle=&-{\rm I}[\hat N,T_\a]T_\ha\eta^{\a\ha}, &&  \langle x_2,\bar J_3 \rangle={\rm I}[\bar J_1,T_\a]T_\ha\eta^{\a\ha}, &  \langle x_3,\bar J_3 \rangle=&{\rm I}[\bar J_2,T_\a]T_\ha\eta^{\a\ha}. \label{xj3b}
\end{align}

Now we show the divergent part of two currents. The first group are the $\langle J_0,\cdot \rangle$ terms:
\begin{align}
 \langle J_0, J_0\rangle=& {\rm I}[J_1,T_\ha][J_3,T_\a]\eta^{\a\ha}-{\rm I}[J_3,T_\a][J_1,T_\ha]\eta^{\a\ha} +{\rm I}[J_2,T_m][J_2,T_n]\eta^{mn},\\
 \langle J_0, J_1\rangle=& -{\rm I}[J_1,T_\ha][N,T_\a]\eta^{\a\ha}-{\rm I}[J_3,T_\a][J_2,T_\ha]\eta^{\a\ha}+{\rm I}[J_2,T_m][J_3,T_n]\eta^{mn},\\
 \langle J_0,\bar J_1\rangle=&- {\rm I}[J_1,T_\ha][\hat N,T_\a]\eta^{\a\ha},\\
 \langle J_0, J_2\rangle=& -{\rm I}[J_3,T_\a][J_3,T_\ha]\eta^{\a\ha}-{\rm I}[J_2,T_m][N,T_n]\eta^{mn},\\
 \langle J_0,\bar J_2\rangle=& {\rm I}[J_1,T_\ha][\bar J_1,T_\a]\eta^{\a\ha}-{\rm I}[J_2,T_m][\hat N,T_n]\eta^{mn},\\
 \langle J_0, J_3\rangle=&{\rm I}[J_3,T_\a][N,T_\ha]\eta^{\a\ha},\\
 \langle J_0,\bar J_3\rangle=&{\rm I}[J_3,T_\a][\hat N,T_\ha]\eta^{\a\ha}+{\rm I}[J_1,T_\ha][\bar J_2,T_\a]\eta^{\a\ha}+{\rm I}[J_2,T_m][\bar J_1,T_n]\eta^{mn}.
\end{align}

The $\langle J_1,\cdot \rangle$ terms are
\begingroup
\allowdisplaybreaks
\begin{align}
 \langle J_1,J_1 \rangle=& -{\rm I}\left([J_2,T_\ha][N,T_\a]-[N,T_\a][J_2,T_\ha]\right)\eta^{\a\ha}+[J_3,T_m][J_3,T_n]\eta^{mn}, \\
 \langle \bar J_1,\bar J_1 \rangle=& 0,\\
 \langle J_1,\bar J_1 \rangle=& \frac{{\rm I}}{2}[\p J_2,T_\ha]T_\a\eta^{\a\ha}+\frac{{\rm I}}{2}\left([J_1,T_i][\bar J_1,T_j]+[\bar J_1,T_i][J_1,T_j]\right)g^{ij} \\
                              & +\frac{{\rm I}}{2}\left(-[\bar J_2,T_\ha][N,T_\a]-[J_2,T\ha][\hat N,T_\a]+3[N,T_\a][\bar J_2,T_\ha]-[\hat N,T_\a][J_2,T_\ha]\right)\eta^{\a\ha},\nonumber\\
 \langle \bar J_1,J_1 \rangle=&-\frac{{\rm I}}{2}[\p J_2,T_\ha]T_\a\eta^{\a\ha}+\frac{{\rm I}}{2}\left([J_1,T_i][\bar J_1,T_j]+[\bar J_1,T_i][J_1,T_j]\right)g^{ij} \\&+\frac{{\rm I}}{2}\left(-[\bar J_2,T_\ha][N,T_\a]-[J_2,T_\ha][\hat N,T_\a]+3[\hat N,T_\a][J_2,T_\a]-[N,T_\a][\bar J_2,T_\ha]\right)\eta^{\a\ha},\nonumber \\
 \langle J_1, J_2 \rangle=& {\rm I}[N,T_\a][J_2,T_\ha]\eta^{\a\ha}+[J_3,T_m][J_3,T_n]\eta^{mn},\\
 \langle \bar J_1, \bar J_2 \rangle=&0,\\
 \langle J_1, \bar J_2 \rangle=&\frac{{\rm I}}{8}[5\p \bar J_3-\bar \p J_3,T_m]T_n\eta^{mn}+\frac{{\rm I}}{8}\left(11[J_2,T_\ha][\bar J_1,T_\a]+5[\bar J_2,T_\ha][J_1,T_\a]\right)\eta^{\a\ha}\nonumber\\&+\frac{{\rm I}}{8}\left(5[\bar J_1,T_i][J_2,T_j]+3[J_1,T_i][\bar J_2,T_j]\right)g^{ij}-\frac{{\rm I}}{2}\left([N,T_a][\bar J_3,T_\ha]+[\hat N,T_\a][J_3,T_\ha]\right)\eta^{\a\ha}\nonumber\\&+\frac{{\rm I}}{2}\left(3[\bar J_3,T_m][N,T_n]-[J_3,T_m][\hat N,T_n]\right)\eta^{mn},\\
 \langle \bar J_1, J_2 \rangle=&-\frac{{\rm I}}{8}[3\p \bar J_3+\bar \p J_3,T_m]T_n\eta^{mn}+\frac{3{\rm I}}{8}\left([J_2,T_\ha][\bar J_1,T_\a]-[\bar J_2,T_\ha][J_1,T_\a]\right)\eta^{\a\ha}\nonumber\\&+\frac{{\rm I}}{8}\left(5[\bar J_1,T_i][J_2,T_j]+3[J_1,T_i][\bar J_2,T_j]\right)g^{ij}-\frac{{\rm I}}{2}\left(3[N,T_a][\bar J_3,T_\ha]-[\hat N,T_\a][J_3,T_\ha]\right)\eta^{\a\ha}\nonumber\\&+\frac{{\rm I}}{2}\left([\bar J_3,T_m][N,T_n]+[J_3,T_m][\hat N,T_n]\right)\eta^{mn},\\
 \langle J_1, J_3 \rangle=&-{\rm I}[N,T_\a][N,T_\ha]\eta^{\a\ha},\label{<j1j3>}\\
 \langle \bar J_1, \bar J_3 \rangle=&-{\rm I}[\hat N,T_\a][\hat N,T_\ha]\eta^{\a\ha},\label{<j1bj3b>}\\
 \langle J_1, \bar J_3 \rangle=&-{\rm I}\left([N,T_\a][\hat N,T_\ha]+[\hat N,T_\a][N,T_\ha]\right)\eta^{\a\ha}+\frac{{\rm I}}{4}\left(3[\bar J_2,T\ha][J_2,T_a]+5[J_2,T_\ha][\bar J_2,T_\a]\right)\eta^{\a\ha}\nonumber\\&+\frac{{\rm I}}{4}\left(5[\bar J_3,T_m][J_1,T_n]+3[J_3,T_m][J_1,T_n]\right)\eta^{mn}+\frac{{\rm I}}{4}\left([J_1,T_i][\bar J_3,T_j]+3[\bar J_1,T_i][J_3,T_j]\right)g^{ij},\\
 \langle \bar J_1, J_3 \rangle=&-{\rm I}\left([N,T_\a][\hat N,T_\ha]+[\hat N,T_\a][N,T_\ha]\right)\eta^{\a\ha}-\frac{{\rm I}}{4}\left([\bar J_2,T\ha][J_2,T_a]-[J_2,T_\ha][\bar J_2,T_\a]\right)\eta^{\a\ha}\nonumber\\&+\frac{{\rm I}}{4}\left([\bar J_3,T_m][J_1,T_n]-[J_3,T_m][J_1,T_n]\right)\eta^{mn}+\frac{{\rm I}}{4}\left([J_1,T_i][\bar J_3,T_j]+3[\bar J_1,T_i][J_3,T_j]\right)g^{ij}.
\end{align}
\endgroup

We present the $\langle J_3,\cdot \rangle$ terms before the $\langle J_2,\cdot \rangle$ due to their similarity with the $\langle J_1,\cdot \rangle$ terms:
\begingroup
\allowdisplaybreaks
\begin{align}
 \langle J_3, J_2 \rangle=&0,\\
 \langle \bar J_3, \bar J_2 \rangle=&-{\rm I}[\hat N,T_\ha][\bar J_1,T_\a]\eta^{\a\ha}-{\rm I}[\bar J_1,T_m][\hat N,T_n]\eta^{mn},\\
 \langle J_3, J_2 \rangle=&\frac{{\rm I}}{8}[5\bar \p J_1-\p \bar J_1,T_m]Tn\eta^{mn}-\frac{{\rm I}}{2}\left([\bar J_1,T_m][N,T_n] - 3[J_1,T_m] [\hat N,T_n]\right)\eta^{mn}\nonumber\\&+\frac{{\rm I}}{2}\left([\hat N,T_\ha][J_1,T_\a]+[N,T_\ha][\bar J_1,T_\a]\right)\eta^{\a\ha}+\frac{{\rm I}}{8}\left(3[\bar J_3,T_i][J_2,T_j]+5[J_3,T_i][\bar J_2,T_j]\right)g^{ij}\nonumber\\&-\frac{{\rm I}}{8}\left(5[J_2,T_\a][\bar J_3,T_\ha]+11[\bar J_2,T_\a][J_3,T_\ha]\right)\eta^{\a\ha},\\
 \langle J_3, J_2 \rangle=&-\frac{{\rm I}}{8}[3\bar \p J_1+\p \bar J_1,T_m]T_n\eta^{mn}+\frac{{\rm I}}{2}\left([\bar J_1,T_m][N,T_n] +[J_1,T_m] [\hat N,T_n]\right)\eta^{mn}\nonumber\\&+\frac{{\rm I}}{2}\left(3[\hat N,T_\ha][J_1,T_\a]-[N,T_\ha][\bar J_1,T_\a]\right)\eta^{\a\ha}+\frac{{\rm I}}{8}\left(3[\bar J_3,T_i][J_2,T_j]+5[J_3,T_i][\bar J_2,T_j]\right)g^{ij}\nonumber\\&+\frac{3{\rm I}}{8}\left([J_2,T_\a][\bar J_3,T_\ha]-[\bar J_2,T_\a][J_3,T_\ha]\right)\eta^{\a\ha},\\
 \langle J_3, J_3\rangle=&0,\\
 \langle \bar J_3, \bar J_3\rangle=&{\rm I}\left([\bar J_2,T_\a][\hat N,T_\ha]-[\hat N,T_\ha][\bar J_2,T_\a]\right)\eta^{\a\ha}+{\rm I}[\bar J_1,T_m][\bar J_1,T_n]\eta^{mn},\\
 \langle \bar J_3, J_3\rangle=&-\frac{{\rm I}}{2}[\bar \p J_2,T_\a]T_\ha\eta^{\a\ha}+\frac{{\rm I}}{2}\left([J_3,T_i][\bar J_3,T_j]+[\bar J_3,T_i][J_3,T_j]\right)g^{ij}\nonumber\\&+\frac{{\rm I}}{2}\left(-[N,T_\ha][\bar J_2,T_\a]-[\hat N,T_\ha][J_2,T_\a]+3[\bar J_2,T_\a][N,T_\ha]-[J_2,T_\a][\hat N,T_\ha]\right)\eta^{\a\ha},\\
 \langle J_3, \bar J_3\rangle=&\frac{{\rm I}}{2}[\bar \p J_2,T_\a]T_\ha\eta^{\a\ha}+\frac{{\rm I}}{2}\left([J_3,T_i][\bar J_3,T_j]+[\bar J_3,T_i][J_3,T_j]\right)g^{ij}\nonumber\\&+\frac{{\rm I}}{2}\left(-3[N,T_\ha][\bar J_2,T_\a]+[\hat N,T_\ha][J_2,T_\a]+[\bar J_2,T_\a][N,T_\ha]+[J_2,T_\a][\hat N,T_\ha]\right)\eta^{\a\ha}.
\end{align}
\endgroup

Finally, the remaining $\langle J_2,\cdot \rangle$ terms:
\begin{align}
 \langle J_2, J_2 \rangle =&{\rm I}[N,T_m][N,T_n]\eta^{mn},\\
 \langle \bar J_2, \bar J_2 \rangle =&{\rm I}[\hat N,T_m][\hat N,T_n]\eta^{mn},\\
 \langle \bar J_2, J_2 \rangle =&-{\rm I}\left([N,T_m][\hat N,T_n]+[\hat N,T_m][N,T_n]\right)\eta^{mn}+\frac{{\rm I}}{2}\left([J_2,T_i][\bar J_2,T_j]+[\bar J_2,T_i][J_2,T_j]\right)g^{ij}\nonumber\\&-\frac{{\rm I}}{2}\left([J_1,T_\a][\bar J_3,T_\ha]-3[\bar J_3,T_\ha][J_1,T_\a] +3[\bar J_1,T_\a][J_3,T_\ha]-[J_3,T_\ha][\bar J_1,T_\a]\right)\eta^{\a\ha},\\
 \langle J_2, \bar J_2 \rangle =&-{\rm I}\left([N,T_m][\hat N,T_n]+[\hat N,T_m][N,T_n]\right)\eta^{mn}+\frac{{\rm I}}{2}\left([J_2,T_i][\bar J_2,T_j]+[\bar J_2,T_i][J_2,T_j]\right)g^{ij}\nonumber\\&+\frac{{\rm I}}{2}\left([\bar J_3,T_\ha][J_1,T_\a]-3[J_1,T_\a][\bar J_3,T_\ha] +3[ J_1,T_\a][\bar J_3,T_\ha]-[\bar J_1,T_\a][J_3,T_\ha]\right)\eta^{\a\ha}.
\end{align}

The terms involving ghost fields that have vanishing anomalous dimension are
\begingroup
\allowdisplaybreaks
\begin{align}
 \langle X, N \rangle=\langle X, \hat N \rangle= \langle \o,\l \rangle = \langle \ho,\hl \rangle=&0, \label{ghost1}\\
 \langle \o, J \rangle=\langle \l ,J\rangle=\langle \ho, \bar J \rangle=\langle \hl ,\bar J\rangle=&0,\\
 \langle \o,\bar J_0 \rangle=\langle \l ,\bar J_0\rangle= \langle \ho,  J_0 \rangle=\langle \hl , J_0\rangle=&0,\\
 \langle \o, N \rangle=\langle \l ,N \rangle= \langle \ho, \hat N\rangle=\langle \hl ,\hat N\rangle=&0,\\
 \langle J, N \rangle=\langle \bar J, \hat N \rangle=&0.
\end{align}
\endgroup

The expressions involving two ghosts and no derivatives are
\begin{align}
 \langle \o,\hl \rangle =&-{\rm I}[\o,T_i][\hl,T_j]g^{ij}, && \langle \l,\ho \rangle =-{\rm I}[\l,T_i][\ho,T_j]g^{ij},\\
 \langle \o,\ho \rangle =&-{\rm I}[\o,T_i][\ho,T_j]g^{ij}, && \langle \l,\hl \rangle =-{\rm I}[\l,T_i][\hl,T_j]g^{ij}.
\end{align}

For one ghost and one current, including the ghost currents,
\begin{align}
 \langle \o, \bar K \rangle=&-{\rm I}[\o,T_i][\bar K,T_j]g^{ij}, &&  \langle \ho, K \rangle=-{\rm I}[\ho,T_i][K,T_j]g^{ij},\\
 \langle \l, \bar K \rangle=&-{\rm I}[\l,T_i][\bar K,T_j]g^{ij} ,&&  \langle \hl, K \rangle=-{\rm I}[\hl,T_i][K,T_j]g^{ij},\\
 \langle \o, \hat N \rangle=&-{\rm I}[\o,T_i][\hat N,T_j]g^{ij}, && \langle \ho, N \rangle=-{\rm I}[\ho,T_i][N,T_j]g^{ij},\\
 \langle \l, \hat N \rangle=&-{\rm I}[\l,T_i][\hat N,T_j]g^{ij}, && \langle \hl, N \rangle=-{\rm I}[\hl,T_i][N,T_j]g^{ij}. \label{ghost2}
\end{align}

Finally, the terms with two currents, with at least one ghost current:
\begin{align}
 \langle \bar K, N \rangle=&-{\rm I}[\bar K,T_i][N,T_j]g^{ij},\\
 \langle K, \hat N \rangle=&-{\rm I}[K,T_i][\hat N,T_j]g^{ij},\\
 \langle N, \hat N \rangle=&-{\rm I}[N,T_i][\hat N,T_j]g^{ij}.
\end{align}

\bibliographystyle{abe}
\bibliography{bibliography}{}

\end{document}